\documentclass[aps,pre,reprint,10pt,twocolumn,showpacs,nofootinbib,superscriptaddress]{revtex4-1}

\usepackage{amsmath}
\usepackage{amsfonts}
\usepackage{amssymb}
\usepackage[pdftex]{graphicx}
\usepackage{subfigure}
\usepackage[pdfborder={0 0 0}]{hyperref}  
\usepackage{color}
\usepackage[latin1]{inputenc}
\usepackage{float}
\usepackage{comment}

\newcommand{\Pde}[1]{{\Pi({#1})}}
\newcommand{\km}{\langle k \rangle}
\newcommand{\ksq}{\langle k^2 \rangle}
\newcommand{\av}[1]{\langle {#1} \rangle}

\begin{document}
\title{Phase transitions with infinitely many absorbing
  states in complex networks}
 
\begin{abstract}
  We instigate the properties of the threshold contact process (TCP),
  a process showing an absorbing-state phase transition with
  infinitely many absorbing states, on random complex networks. The
  finite size scaling exponents characterizing the transition are
  obtained in a heterogeneous mean field (HMF) approximation and
  compared with extensive simulations, particularly in the case of
  heterogeneous scale-free networks. We observe that the TCP exhibits
  the same critical properties as the contact process (CP), which
  undergoes an absorbing-state phase transition to a single absorbing
  state. The accordance among the critical exponents of different
  models and networks leads to conjecture that the critical behavior
  of the contact process in a HMF theory is a universal feature of
  absorbing state phase transitions in complex networks, depending
  only on the locality of the interactions and independent of the
  number of absorbing states. The conditions for the applicability of
  the conjecture are discussed considering a parallel with the
  susceptible-infected-susceptible epidemic spreading model, which in
  fact belongs to a different universality class in complex networks.
\end{abstract}

\author{Renan S. Sander}

\author{Silvio C. Ferreira} 
\email{silviojr@ufv.br}

\affiliation{Departamento de F\'{\i}sica, Universidade Federal de
  Vi\c{c}osa, 36571-000, Vi\c{c}osa - MG, Brazil} 

\author{Romualdo Pastor-Satorras} 

\affiliation{Departament de F\'{\i}sica i Enginyeria Nuclear,
  Universitat Polit\`ecnica de Catalunya, Campus Nord B4, 08034
  Barcelona, Spain}


\maketitle

\section{Introduction}

\label{sec:intro}

The role of the effects of a disordered substrate on the behavior of
dynamical processes has attracted the attention of the statistical
physics community since long
\cite{hugues95:_random_walks_II,havlindiffusiondisorder2000}. The
interest in this issue has been enhanced by the recent realization
that the substrate of many relevant dynamical processes can be
represented in terms of a complex network
\cite{barabasi02,mendesbook,newman2010}. This realization has led to a
renewed scientific effort emphasizing the effects of the disordered
topology inherent to the underlying network
\cite{dorogovtsevRMP08,barratbook}. In this context, it has been shown
that highly heterogeneous contact patterns, characterized by a degree
distribution $P(k)$ (probability that an element or vertex in the
network is connected to other $k$ vertices) showing a heavy-tailed
form
% with a diverging second moment $\av{k^2}$
can have a very strong impact in processes such as epidemic spreading
\cite{pv01a,May01,PhysRevE.66.016128}, percolation
\cite{Cohen2000,PhysRevLett.85.5468}, or synchronization
\cite{ArenasPhysRep2008}.

Among the different dynamical processes considered in networks, an
important role has been played by the class of processes with
absorbing states~\cite{marro1999npt,Henkel}. Absorbing states are
configurations of the system in which the dynamics becomes trapped
forever and that usually imply the presence of absorbing state phase
transitions (APTs) between an active phase and the absorbing
configurations. The simplest model exhibiting an APT is the contact
process (CP)~\cite{harris74}, where particles lying on a lattice or
network undergo spontaneous annihilation and catalytic creation.  A
creation event involves a pair of empty and occupied nodes and is
controlled by a rate $\lambda$.  The CP undergoes a continuous APT at
a critical point $\lambda_c$, separating an absorbing phase with all
sites empty from an active one with a nonzero order parameter $\phi$,
defined as the average density of particles at the steady state.  The
ensuing APT is characterized in terms of a set of critical exponents,
that define the so-called directed percolation (DP) universality class
\cite{Henkel}.  The theoretical analysis of the CP in
networks~\cite{RomuPRL2008,bogunaPRE2009,Ferreira_annealed}, based in
the heterogeneous mean-field (HMF) theory
\cite{dorogovtsevRMP08,barratbook}, has shown the presence of an APT
in which the scaling of the relevant critical quantities with network
size presents very strong corrections to scaling in scale-free (SF)
networks with an heterogeneous degree distribution given by a power
law form, $P(k) \sim k^{-\gamma}$~\cite{barabasi02}. These corrections
to scaling, albeit very large, have been observed in large-scale
numerical simulations~\cite{Ferreira_quenched}.

While simple models such as the CP are characterized by a unique
absorbing state devoid of all particles, other relevant systems
present multiple absorbing configurations~\cite{WIJLAND2003}. Examples
of those range from models of forest fires~\cite{Drossel92} to
self-organized critical sandpiles~\cite{PhysRevE.57.5095}. A basic
model presenting an APT with multiple absorbing states is the pair
contact process (PCP), where a pair of occupied neighbors annihilates
with probability $p$ and creates a new particle in one nearest
neighbor (NN) of the pair with probability
$1-p$~\cite{JensenPCP93}. Any configuration with only isolated
particles is an absorbing state, which implies that the PCP has
infinitely many absorbing configurations in the thermodynamic
limit. On lattice, the PCP belongs to the DP universality
class~\cite{Henkel}.

In the field of complex networks, dynamical systems with many
absorbing states have been used to investigate self-organized
criticality and avalanches~\cite{Goh03,Andrade10,Lee04}, while the
analysis of the ensuing APT is limited to a few
works~\cite{Bancal10,da-yin10}. The HMF analysis of the forest fire
model in heterogeneous networks yielded critical exponents equal to
those obtained for the susceptible-infected-susceptible epidemic model
in the same approach~\cite{Bancal10}.  A finite-size scaling
\cite{cardy88} analysis of the PCP was performed for homogeneous
networks~\cite{da-yin10}, but using very limited sizes, which do not
allow to extract reliable conclusions about the model's universality
class.

In this work, we present and analyze a model possessing an APT with
many absorbing states, which is amenable to analytical calculations
based on the HMF formalism and large-scale numerical simulations.  The
model, that we call the threshold contact process (TCP), is inspired
in the PCP, with the peculiarity that the \textit{fermionic}
constraint of single occupation of vertices is relaxed to allow double
occupation. This modification hugely reduces the computer algorithmic
complexity and allows us to perform a standard HMF
analysis~\cite{dorogovtsevRMP08,barratbook,Vespignani12}. We show that
the behavior of the TCP at the HMF level is the same of the CP, a fact
that we confirm by means of extensive numerical simulations based in
the quasi-stationary state method
\cite{Mancebo2005,Ferreira_annealed,Ferreira_quenched}. We thus
conclude that, despite the non-fermionic nature of the TCP, it extends
to complex networks the fundamental inclusion of the PCP in the
universality class of the contact process.

We have organized our paper as follows: In
Sec.~\ref{sec:cont-proc-netw} we present a brief overview of the
properties of the CP on complex networks. Section~\ref{sec:model} is
devoted to describe the TCP model. A HMF theory and a homogeneous
pair-approximation for TCP are developed in Sec.~\ref{sec:mf}. The
simulation methods and numerical procedures used are described in
Sec.~\ref{sec:simu}, where we also present a check of the HMF theory
by means of numerical simulations on annealed networks.  Simulations
on real quenched networks are presented in Sec.~\ref{sec:quenched},
where we compare them with the HMF theory predictions.  Finally, we
draw our concluding remarks in Sec.~\ref{sec:conclu}.

\section{The contact process on networks}
\label{sec:cont-proc-netw}

The CP dynamics on networks is defined as follows~\cite{RomuPRL2006}:
Every node can be either occupied or empty. Occupied nodes become
empty at a rate $1$, while empty sites become occupied at rate
$\lambda n /k$, where $k$ is the number of NNs of the node (the node's
degree), and $n$ the number of occupied nodes among the NNs.

The properties of the CP dynamics on heterogeneous networks have been
analytically investigated within the framework of the heterogeneous
mean-field (HMF) theory
\cite{RomuPRL2006,RomuPRL2008,bogunaPRE2009,Ferreira_annealed}, an
extension of the standard mean-field theory taking into account the
network heterogeneity.  The HMF theory assumes that the vertex degree
is the only relevant quantity, substitutes the topological structure
of the network by an annealed form~\cite{dorogovtsevRMP08,barratbook},
and neglects all dynamical correlations between vertices.
%This last assumption is expected
%\textit{a priori} to be correct due to the small-world property of
%most real networks~\cite{watts98}. 
In HMF theory, focus is placed on
the partial densities of particles in vertices of degree $k$, denoted
by $\phi_k$, which satisfy the rate equation~\cite{RomuPRL2006}
\begin{equation}
  \label{eq:1}
  \frac{d \phi_{k}}{dt}=-\phi_{k}+\lambda
  k[1-\phi_{k}]\sum_{k'}\frac{P(k'|k)\phi_{k'}}{k'}.
\end{equation}
The quantity $P(k'|k)$ is a measure of the degree correlations
\cite{mendesbook}, defined as the conditional probability that a
vertex of degree $k$ is connected to a vertex of degree $k'$
\cite{alexei}.  The solution of this equation in the stationary limit
leads to an APT located at the critical point $\lambda_c=1$,
irrespective of the degree correlations in the network. For degree
uncorrelated networks, in which $P(k'|k) = k' P(k')/\av{k}$
\cite{mendesbook}, a complete solution can be worked out
\cite{RomuPRL2006}, giving an order parameter, defined as the average
particle density, that close to the critical point scales as $\phi =
\sum_k P(k) \phi_k \sim (\lambda-\lambda_c)^\beta$, a characteristic
time scale $\tau \sim (\lambda-\lambda_c)^{-\nu}$ and a decay of the
particle density at the critical point $\phi_c(t) \sim
t^{-\delta}$. In SF networks with degree distribution $P(k) \sim
k^{-\gamma}$, one has $\beta = \delta=\max\lbrace 1/(\gamma-2),1\rbrace$,
while $\nu=1$ for any $\gamma$.

These critical exponents, depending on the degree distribution, are
valid in the thermodynamic limit of an infinite network.  
In finite systems, however, the transition to the absorbing state is
better characterized in terms of a finite-size scaling (FSS) analysis
\cite{cardy88}.
In Refs.~\cite{RomuPRL2008,NohPRE2009,Ferreira_annealed} such theory
has been developed, exploiting a mapping of the CP in the low density
regime to a one-step process. Within this approximation, one can build
the equation for the probability distribution $\bar{P}_{n}$ of
observing $n$ active particles in the steady state,
which in the thermodynamic limits takes a scaling form given
by~\cite{Ferreira_annealed} 
\begin{equation}
 \label{eq:pnqs}
 \bar{P}_n = \frac{1}{\sqrt{\Omega}}f\left(\frac{n}{\sqrt{\Omega}}\right),
\end{equation}
where $f$ is a scaling function normalized as $\int_0^\infty f(x)dx
=1$, $\Omega=N/g$ and $g=\langle k ^2\rangle/\langle k \rangle^2$.
The critical density and characteristic time follow directly from this
distribution~\cite{Ferreira_annealed} and are given by
\begin{equation}
  \bar{\phi}\sim (N g)^{-1/2}~~~~\mbox{and}~~~~\tau\sim (N/g)^{1/2}.
  \label{eq:FSS}
\end{equation}
For a power law degree-distribution, the factor $g$ diverges with the
degree cutoff $k_c$ (the largest degree present in the network) 
as $g\sim k_c^{3-\gamma}$ for $\gamma<3$ and
becomes constant (logarithmic) for $\gamma>3$
($\gamma=3$)~\cite{mariancutofss}. For a general cutoff scaling as
$k_c\sim N^{1/\omega}$, the scaling laws 
\begin{equation}
  \label{eq:4}
  \bar{\phi} \sim N^{-\hat{\nu}}~~~~\mbox{and}~~~~\tau\sim
  N^{\hat{\alpha}} 
\end{equation}
ensue in the limit $N\to\infty$, with associated critical exponents
\cite{Ferreira_annealed} 
\begin{equation}
 \hat{\nu} =
 \frac{1}{2}+\max\left(\frac{3-\gamma}{2\omega},0\right), \;
 \hat{\alpha}=\frac{1}{2}-\max\left(\frac{3-\gamma}{2\omega},0\right).
 \label{eq:nu_alpha}
\end{equation}
This exponents, which define what we can call the CP universality
class at the HMF level (the CP-HMF class), have been proved to provide
a correct description of the CP in networks by means large-scale
numerical simulations~\cite{Ferreira_quenched}, corroborating thus the
validity of the description of the CP on networks in terms of the HMF
theory.

\section{The threshold contact process}
\label{sec:model}

One of the simplest models exhibiting a transition to infinitely many
absorbing states is the PCP~\cite{JensenPCP93}. An optimized computer
simulation of PCP near the critical point (low density of active
sites) requires the maintenance of an updated list containing the
coordinates of all active pairs. This task becomes computationally
inefficient when the model is implemented on an arbitrary graph, more
specifically during the annihilation events when the involved pair and
all the corresponding active links must be removed from the list.
Kockelkoren and Chat\'e~\cite{Kockelkoren} had already realized that
the fermionic constraint can be counterproductive in lattices due to
algorithmic complexity and, more importantly, due to strong deviations
to scaling originated from this restriction.  For this reason, we
consider a modification of the PCP, that we call the threshold contact
process (TCP), that is defined by the following rules: Nodes can be
occupied by 0, 1 or 2 particles (i.e. they can have an occupation
number $\sigma_i=0,~1,~2$). Particles in a doubly occupied node can be
both annihilated at rate 1 or either create an offspring particle in a
randomly chosen NN at rate $\lambda$. The creation event takes
actually place only if the selected NN is empty or singly
occupied. Because empty and singly occupied nodes do not react, any
configuration devoid from doubly occupied sites is an absorbing
state. Singly occupied sites do however play a fundamental role, since
they form a backbone of connected vertices through which activity can
spread.  The TCP can be easily generalized to higher thresholds
(maximum occupation number), $h>2$.

The computer implementation of TCP in an arbitrary graph can be done
with the following scheme: At each time step, an active (doubly
occupied) vertex $j$ is randomly chosen and time is updated as
$t\rightarrow t+\Delta t$, where $\Delta t = 1/[(1+\lambda)n(t)]$ and
$n(t)$ is the number of active vertices at time $t$. With probability
$p=1/(1+\lambda)$, both particles of the selected vertex are
annihilated, producing an empty vertex. With the complementary
probability $1-p=\lambda/(1+\lambda)$, one of the $k$ neighbors of $j$
is randomly selected and, if empty or singly occupied, receives a new
particle; otherwise nothing happens and the simulation proceeds to the
next time step.

{While we have checked numerically that the TCP belongs to 
the DP class in regular $d$-dimensional lattices (to be reported elsewhere), 
in the following we will consider its behavior in complex networks.}

\section{Mean-field analysis}
\label{sec:mf}

\subsection{Homogeneous mean-field theory}
\label{sec:hom}

In homogeneous substrates such as a regular lattice, all nodes are
equivalent and thus the equations of motion are the same for all of
them. Letting $\Pi(\mathbf{\sigma}_1,\cdots,\sigma_l)$ be the
probability that a cluster of $l$ vertices has the configuration
$\lbrace\sigma_1\cdots\sigma_l\rbrace$, we can find the following
exact equations of motion for single sites in the TCP dynamics:
\begin{eqnarray}
 \frac{d\Pde{2}}{dt} &=& -\Pde{2}+\lambda\Pde{1,2}
\label{eq:homo1}\\
 \frac{d\Pde{1}}{dt} &=& \lambda[\Pde{0,2}-\Pde{1,2}].
\label{eq:homo2}
\end{eqnarray}
The equations for single nodes depend on pairs, those for pairs depend
on triplets (as we will see below), and so forth.  An approximate
solution can be obtained by breaking this hierarchy of equations by
means of some closure condition.  Thus, in the simplest one-site
approximation, the correlation between pairs is neglected, resulting
in $\Pde{\sigma,\sigma'}\approx
\Pde{\sigma}\Pde{\sigma'}$~\cite{marro1999npt}.  Equations
(\ref{eq:homo1}) and (\ref{eq:homo2}) in this one-site approximation
become
\begin{eqnarray}
 \frac{d\phi}{dt} &=& -\phi+\lambda\rho\phi,
\label{eq:mf_phi}\\
\frac{d\rho}{dt} &=& \lambda\phi(1-\phi-2\rho),
\label{eq:mf_rho}
\end{eqnarray}
where we have defined $\phi=\Pde{2}$, $\rho=\Pde{1}$, and obviously
$\Pi(0) = 1-\rho-\phi$. In the stationary regime, Eqs.~(\ref{eq:mf_phi})
and~(\ref{eq:mf_rho}) yield $\rho=1/\lambda$ and $\phi =
(\lambda-2)/\lambda$. We can therefore identify the critical point
$\lambda_c=2$ and the DP mean field exponent
$\beta=1$~\cite{marro1999npt}. One can easily verify that the
characteristic time $\tau$ scales with DP mean field exponent, $\tau
\sim (\lambda-\lambda_c)^{-1}$, while the density of active sites
decays with time at criticality as $\phi_c\sim t^{-1}$.

For the case of the CP, it has been shown that a homogeneous pair
approximation (HPA)~\cite{Avraham92}, considering the hierarchy of
equations up to second order, yields better estimates of the critical
point even in the realistic case of heterogeneous
networks~\cite{MunozPRL2010,Ferreira_quenched}. We can also follow
this HPA approach for the TCP. Firstly, we observe that in the
stationary regime, Eqs.~(\ref{eq:homo1}) and (\ref{eq:homo2}) give
$\Pde{2,0}=\Pde{2,1} = {\phi}/{\lambda}$. Moreover, using the
normalization condition $\Pde{0,2}+\Pde{1,2}+\Pde{2,2}=\Pde{2}$, we
have $\Pde{2,2} = 1-2\phi/\lambda$.  The equations for the remaining
independent pairs are
\begin{eqnarray}
 \frac{d\Pde{0,0}}{dt} &=& 
 2 \Pde{2,0}- 2\frac{q-1}{q}\lambda\Pde{0,0,2},
 \label{eq:00}\\
 \frac{d\Pde{0,1}}{dt} &=& 
 \lambda\frac{q-1}{q}[\Pde{0,0,2}-\Pde{0,1,2}],
\label{eq:01}\\
 \frac{d\Pde{1,1}}{dt} &=&  
2\lambda\frac{q-1}{q} [\Pde{1,0,2}-\Pde{1,1,2}].
\label{eq:11}
\end{eqnarray}
Here $q$ is number of NNs of a vertex, which we assume to be constant
for all vertices in this homogeneous approach.  Furthermore, since
homogeneity implies $\Pde{\sigma,\sigma'} = \Pde{\sigma',\sigma}$, the
evolution of all pair is determined given Eqs.~\eqref{eq:00}-\eqref{eq:11}.

Up to this point, all pair motion equations are exact. Now, we perform
the pair approximation~\cite{Avraham92}
\begin{equation}
 \Pde{\sigma,\sigma',\sigma''}\approx 
 \frac{\Pde{\sigma,\sigma'}\Pde{\sigma',\sigma''}}{\Pde{\sigma'}},
 \label{eq:pair}
\end{equation}
which, after applying normalization and symmetry conditions, leads to
the following non-linear equations for the stationary state:
\begin{eqnarray}
  \rho -\frac{\phi}{\lambda} &=&\frac{q}{\lambda(q-1)}
  \frac{\rho(1-\phi)}{1-\rho-\phi} 
 \label{eq:trans1}\\
 \rho &=& 1 -\frac{q}{\lambda(q-1)}
-\frac{\phi}{\lambda}\left[\lambda+1-\frac{q}{q-1}\right].
 \label{eq:trans2}
\end{eqnarray}
Solving Eqs.(\ref{eq:trans1}) and (\ref{eq:trans2}) we obtain
\begin{equation}
 \phi = \frac{\lambda -\lambda_c(q)}{\lambda-\lambda_c(q)+2}, 
\end{equation}
where the critical point for general $q$ is given by
\begin{equation}
\lambda_c(q) = \frac{2q}{q-1}.
\label{eq:lbc_pair}
\end{equation}
In Sec.~\ref{sec:simu} we will show that this result fits well the
critical point obtained from computer simulations of TCP on networks.

\subsection{Heterogeneous mean field theory}
\label{sec:hmf}

Application of HMF to the TCP in heterogeneous networks follows the
steps performed for the CP. In the heterogeneous case, we
differentiate the functions $\rho$ and $\phi$ according to degree,
defining $\phi_k$ and $\rho_k$ as the probability that a vertex of
degree $k$ is doubly or simply occupied, respectively.  The rate
equations (\ref{eq:mf_phi}) and (\ref{eq:mf_rho}) are thus simply
modified within the approximation, taking now the form
\begin{eqnarray}
\label{eq:dphik}
\frac{d\phi_k}{dt} &=& -\phi_k+\lambda k \rho_k  \Theta_k \\
\label{eq:drhok}
\frac{d\rho_k}{dt} &=& \lambda k (1-2\rho_k-\phi_k)\Theta_k,
\end{eqnarray}
where
\begin{equation}
\label{eq:Thetak}
\Theta_k = \sum_{k'}\frac{P(k'|k)\phi_{k'}}{k'},
\end{equation}
is the probability that a vertex of degree $k$ receives a particle
from an active NN. As in the CP, the factor $1/k'$ accounts for the
random choice of the target neighbor for offspring creation from a
vertex of degree $k'$. In the case of degree uncorrelated networks, we
are led to the simplified coupling factor $\Theta_k \equiv \Theta=
\sum_{k} \phi_k P(k)/\av{k} \equiv \phi/\av{k}$, independent of $k$.

In order to solve Eqs.~\eqref{eq:dphik} and \eqref{eq:drhok}, we
notice that $\phi_k$ gives the order parameter $\phi=\sum_kP(k)\phi_k$
in the APT. The density $\rho=\sum_kP(k)\rho_k$ is a non-critical
field coupled to the order parameter that goes to a finite value at
the steady state even at the critical point or in the subcritical
phase. We thus assume that it will reach its steady state in a way
faster than $\phi$. We have checked that this assumption is compatible
with our numerical simulations (see Fig.~\ref{fig:rlx}). We thus
proceed to perform a quasi-static approximation
\cite{michelediffusion}, imposing $\dot{\rho}_k \simeq 0$ in
Eq.~(\ref{eq:drhok}) to obtain
\begin{equation}
\rho_k \simeq \frac{1-\phi_k}{2},
\label{eq:drhok2}
\end{equation}
which, upon substitution into Eq.~\eqref{eq:dphik}, leads to the equation
\begin{equation}
\label{eq:dphik2}
\frac{d\phi_k}{dt} \simeq  -\phi_k+\frac{\lambda}{2} k (1-\phi_k)  \Theta_k,
\end{equation}
that holds for long times.

Equation (\ref{eq:dphik2}) is exactly the same HMF rate equation
obtained for the order parameter of the CP on heterogeneous networks
(see Eq.~\eqref{eq:1}), with an effective creation rate $\lambda' =
\lambda/2$. Therefore, we can exploit all results presented in
Sec.~\ref{sec:cont-proc-netw}: For an infinite network size, the TCP
will undergo a continuous APT at a critical $\lambda_c=2$,
independently of the connectivity pattern of the network. Under the
additional simplification of degree uncorrelated networks, we recover
the critical behavior $\phi\sim(\lambda-\lambda_c)^\beta$ with $\beta
= \max\lbrace 1/(\gamma-2),1\rbrace$, while the characteristic time
diverges as $\tau\sim |\lambda-\lambda_c|^{-1 }$.  At the critical
point, the order parameter decays as $\phi\sim t^{-\delta}$, with
$\delta=\beta$.  In finite networks, on the other hand, the same
mapping to a one-step process performed for the CP ensues from
Eq.~\eqref{eq:dphik2} (with the corresponding mapping $\lambda' =
\lambda/2$), and therefore we expect for the TCP a FSS form given by
Eqs.~\eqref{eq:FSS}.

We finally note that Eqs.~(\ref{eq:drhok2}) and (\ref{eq:dphik2}) lead
to a stationary density of singly occupied nodes given by
\begin{equation}
% \phi_k = \frac{\lambda k\phi}{2\langle k \rangle}+\mathcal{O}(2), 
 %\label{eq:phik}%\\
 \rho_k = \frac{1}{2}-\frac{\lambda k\phi}{4\langle k
   \rangle}+\mathcal{O}[(k\phi)^2]. 
 \label{eq:rhok}
\end{equation}

Therefore, at criticality about 50\% of the nodes belong to the
backbone of singly occupied vertices through which dynamic
spreads. This result is verified with good accuracy in our simulation
results, presented in Sec.~\ref{sec:quenched}.

\section{Simulation methods}
\label{sec:simu}

\subsection{The quasi-stationary state method}
\label{sec:quasi-stat-state}

In infinite systems, APTs are sharply defined. In finite systems,
however, the absorbing state can be visited even in the supercritical
phase due to stochastic fluctuations. The numerical exploration of
APTs in this case is a subtle problem, that must be handled with
suitable simulation strategies. Here we investigate the stationary
properties of the TCP using the quasi-stationary (QS)
method~\cite{Mancebo2005}, which has been successfully applied to
determine with high accuracy the critical properties of the CP in SF
networks~\cite{Ferreira_annealed,Ferreira_quenched}. In the QS method,
every time the system tries to visit an absorbing configuration, the
current configuration is replaced by one active state picked up
randomly from the system history. The method is implemented by storing
and constantly updating a list containing $M$ active configurations
previously visited by the dynamics. An update consists in randomly
selecting a configuration in the list and replacing it by the present
active configuration with a probability $p_{r}\Delta t$. After a
relaxation time $t_r$, the QS probability $\bar{P}_n$, corresponding
to states with $n$ occupied vertices, is computed during an averaging
interval $t_a$. All relevant QS quantities can be then determined from
$\bar{P}_n$ as, for example, the density of active particles
$\bar{\phi} = N^{-1}\sum_n n\bar{P}_n $ and the characteristic
lifespan $\tau =1/\bar{P}_1$~\cite{Ferreira_annealed}.

The networks used for simulations were generated using the
uncorrelated configuration model (UCM)~\cite{Catanzaro05}, with a
minimum degree $m=6$ and a hard cutoff $k_c=N^{1/2}$, which leads to
the absence of the degree-degree correlations in large quenched SF
networks~\cite{mariancutofss} without self nor multiple connections.
System sizes with up $10^7$ nodes were used.  Simulations were
performed with fixed parameters $t_r=4\times10^6$, $t_a = 4 \times
10^7$ and $M=200$.  The list with the system history must be updated
according to the frequency that the system tries to visit the
absorbing state. The larger the lifespan $\tau$ the slower the list
must be updated in the stationary state. In our simulations, we used
$p_r=10^{-3} - 10^{-2}$, being the larger values used for the smaller
network sizes.

\subsection{Critical point determination: The moment ratio method}
\label{sec:crit-point-determ}

A key point in the numerical computation of critical exponents in any
phase transition is the accurate determination of the critical point.
The usual strategy to determine critical points in lattice models
considers the search for pure power laws in plots $\bar{\phi}$ or
$\tau$ versus $N$~\cite{marro1999npt}. This strategy is however not
well suited in the presence of strong corrections to scaling, as
expected for the TCP, given its relation with the CP. In
Ref.~\cite{Ferreira_quenched}, it was proposed an alternative method,
using the idea of size-independent critical moment
ratios~\cite{Landaubook} in the form
\begin{equation}
M^{n}_{qs}=\frac{\langle\phi^n\rangle}{\langle\phi^q\rangle
\langle\phi^s\rangle},~~~~q+s=n.
\label{eq:moments}
\end{equation}
From the scaling form of $\bar{P}_n$, Eq.~(\ref{eq:pnqs}), we have
\begin{equation}
  \label{eq:2}
  \langle n^k\rangle =  \sum_{n=1}^{N}\frac{n^k}{\sqrt{\Omega}} f\left(
    \frac{n}{\sqrt{\Omega}} \right).
\end{equation}
Defining $x=n/\sqrt{\Omega}$ with increments $\Delta x = 1/\sqrt{\Omega}$ we
find
\begin{equation}
  \label{eq:3}
  \langle {n^k} \rangle = \sqrt[k]{\Omega}\sum_{x=1/\sqrt{\Omega}}^{N/\Omega}
  x^k f(x)\Delta x 
~\stackrel{N\rightarrow \infty}{=}~a_k \sqrt[k]{\Omega}
\end{equation}
where $a_k = \int_0^\infty x^kf(x)dx$. Therefore, the asymptotic
moments ratios at mean field level are size-independent and given by
$M^{n}_{qs} = {a_n}/{a_q a_s}$. One can thus estimate the critical
point taking advantage of this fact: Plotting the quantities
$M_{n}^{qs}$ as a function of $\lambda$ for different values of $N$,
the different curves will intersect at the critical point
$\lambda=\lambda_c$, for all different network
sizes~\cite{Landaubook}. Obviously, the validity of this scheme is
conditioned to the scaling form of $\bar{P}_n$, that holds only for
large systems, even at a mean field level~\cite{Ferreira_annealed}.

\subsection{Check of numerical methods}
\label{sec:ann}

In order to check the efficiency of our simulation and numerical
methods, we have considered the TCP on top of annealed networks
\cite{stauffer_annealed2005,bogunaPRE2009}, in which all edges are
rewired between any two dynamical steps, while preserving the degree
and degree correlations of each individual node. This procedure
destroys all dynamical correlations and thus renders exact the
predictions of HMF theory \cite{dorogovtsevRMP08,bogunaPRE2009}. In
practice, simulations on annealed networks are carried out in
uncorrelated networks by selecting at random, every time that a node
has to choose a nearest neighbor, a vertex of degree $k$ in the whole
network with a probability $k P(k)/\langle
k\rangle$~\cite{bogunaPRE2009}. In the following, we consider SF
networks with a degree distribution $P(k) \sim k^{-\gamma}$ and
$\gamma>2$.

The HMF theory discussed in section~\ref{sec:hmf} predicts a FSS
explicitly depending on the factor $g=\langle k^2 \rangle/\langle k
\rangle^2$ that, for SF networks with $2<\gamma<3$, behaves as $g
\simeq \mathrm{const} \times \left[1 - \xi^{3-\gamma} +
  2\xi^{\gamma-2} \cdots\right] k_c^{3-\gamma}$, where
$\xi=k_0/k_c$. So, even in annealed networks the FSS is ruled by
strong corrections, specially for $\gamma\approx 3$ and $\gamma\approx
2$ when they become logarithmic. However, since HMF theory is exact in
this case, the critical point must be $\lambda_c=2$. We have checked
that the moment ratio method precisely recovers this theoretical
expectation, with all moment ratios, for different values of $q$ and
$s$ and different network sizes, intersecting at $\lambda_c=2$ for any
value of $\gamma$.

\begin{figure}[t]
\centering
 \includegraphics[width=8cm]{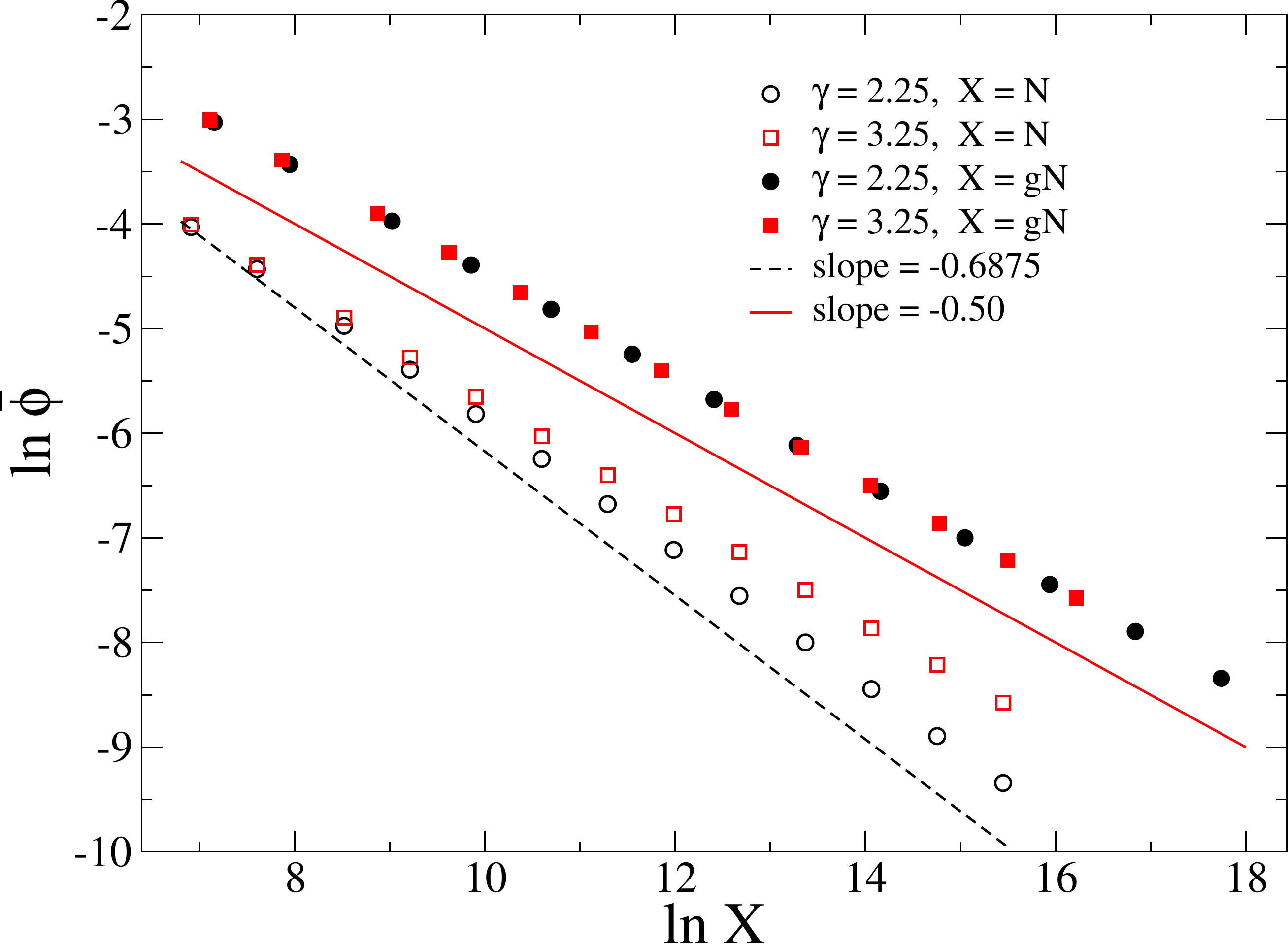}\\~\\
 \includegraphics[width=8cm]{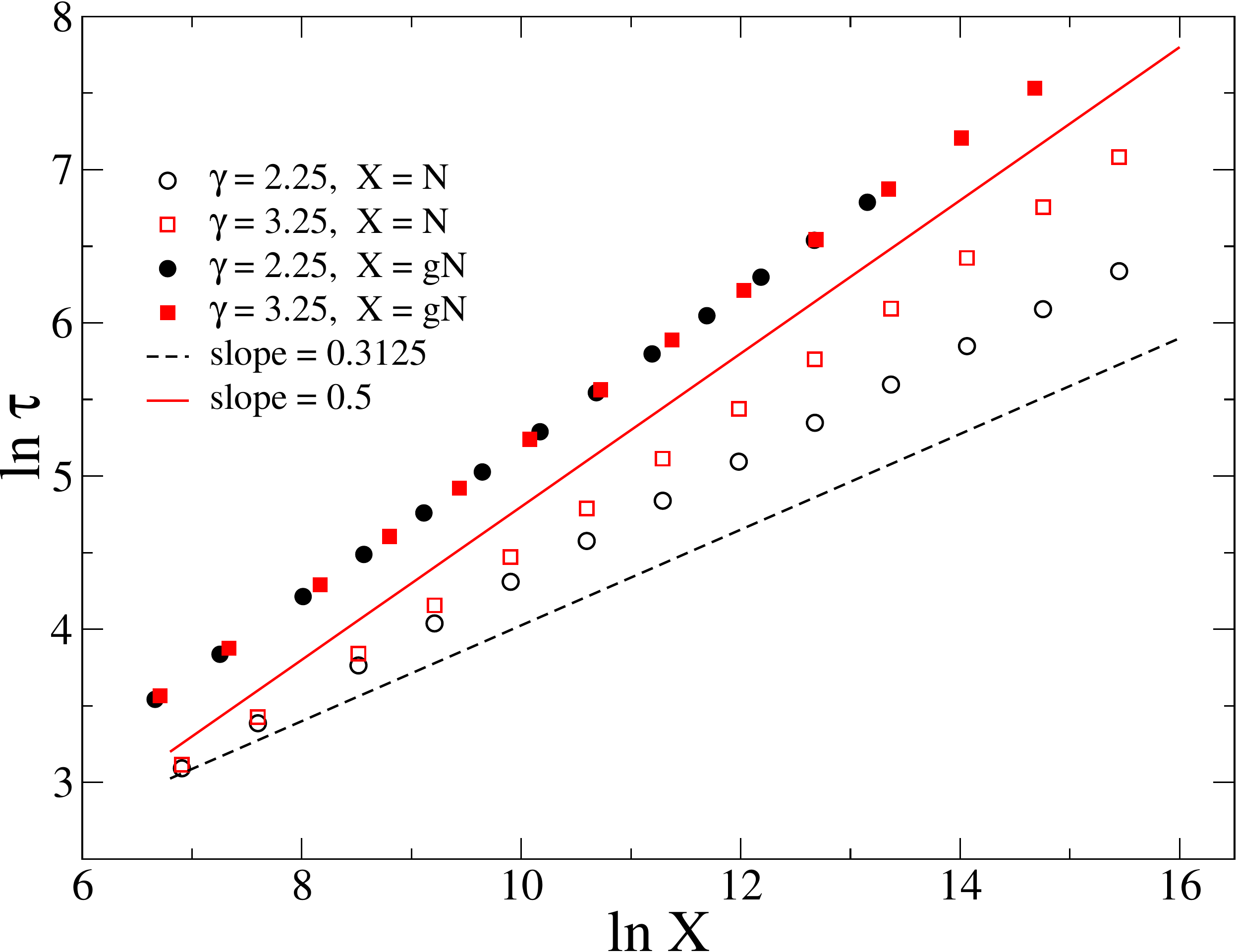}
 \caption{Finite size scaling of the critical density (top) and
   characteristic lifespan (bottom) for the TCP on annealed SF
   networks. Symbols represent simulations; dashed lines are the
   scaling as a function of $N$ expected from Eq.~\eqref{eq:nu_alpha}
   for $\gamma=2.25$; full lines are the HMF theory predictions for
   $\gamma>3$, Eq.~\eqref{eq:nu_alpha}, or for all $\gamma$ when
   taking into account the full $g$ dependence,
   Eq.~\eqref{eq:FSS}. Filled symbols were shifted upwards to improve
   visibility. }
 \label{fig:QSannealed}
\end{figure}

Figure~\ref{fig:QSannealed} shows the QS density and characteristic
lifespan for critical TCP on annealed networks with degree exponents
$\gamma=2.25$ and $\gamma=3.25$. The dependence of these quantities
with the system size neatly deviates from the pure power law at
criticality given by Eq.~\eqref{eq:nu_alpha}. This is however, a
simple finite size effect; when the full factor $g$ is explicitly
included, we obtain a perfect agreement with HMF exponents
$\bar{\phi}\sim (gN)^{-S_\nu}$ and $\tau\sim(N/g)^{S_\alpha}$ with
$S_\nu=S_\alpha=1/2$. The exponents obtained by means of linear
regressions are $S_{\nu} = 0.498(3)$ and $0.502(3)$ while $S_\alpha =
0.499(2)$ and $0.498(2)$ for $\gamma = 2.25$ and $3.25$, respectively.

%\section{TCP on real quenched networks}
\section{TCP on quenched networks}
\label{sec:quenched}

\begin{figure}[t]
  \centering
  \includegraphics[width=8cm]{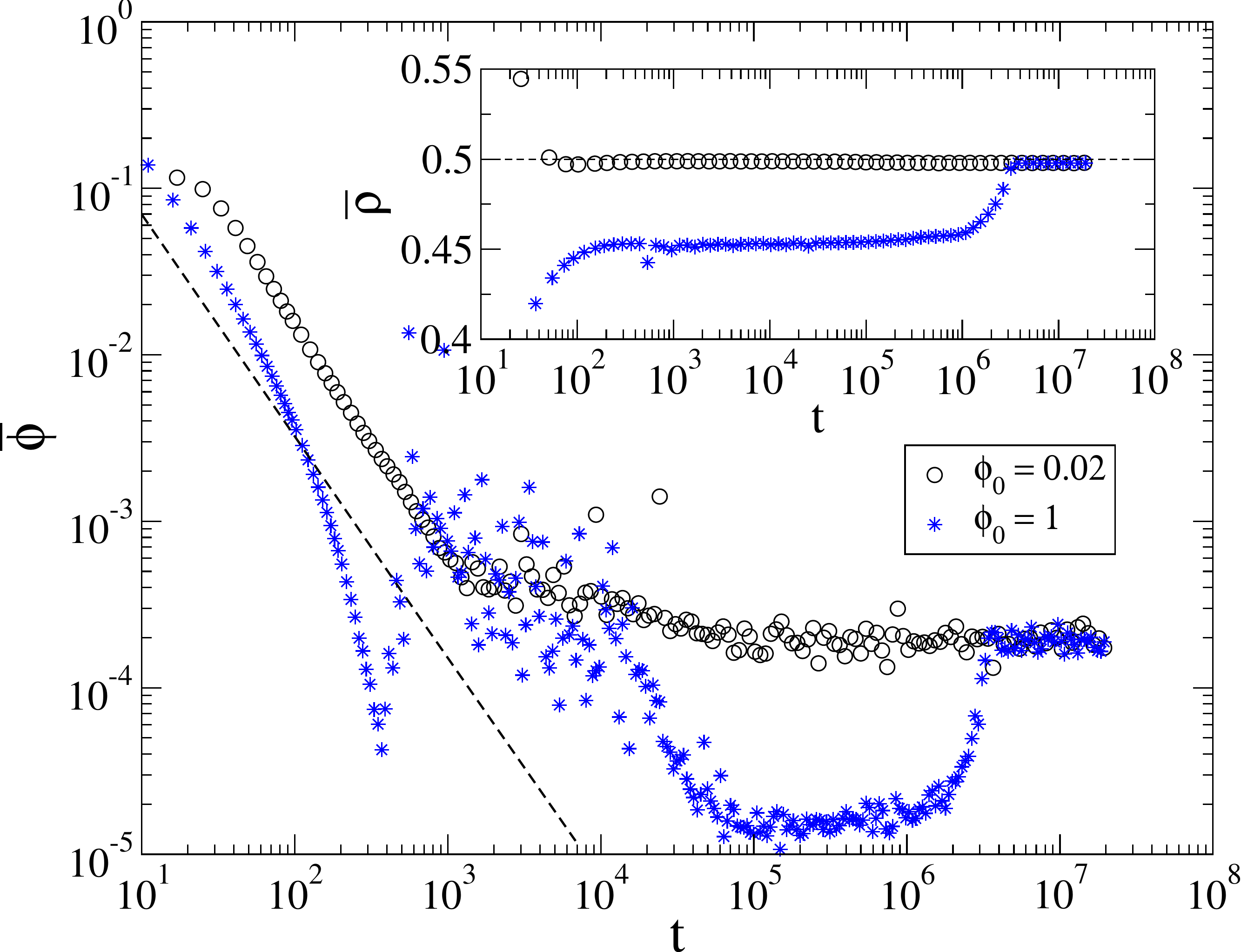}
  % phivstg275N640k.pdf: 764x565 pixel, 72dpi, 26.95x19.93 cm, bb=0 0 764 565
  \caption{Critical relaxation of the density of active vertices for
    QS simulation of TCP on quenched SF networks with $\gamma=2.75$
    and size $N=1.28\times 10^6$ Two initial conditions are reported:
    Fully active network (crosses) and $2\%$ of randomly distributed
    active vertices (circles).  The dashed line is the power law decay
    $\phi\sim t^{-4/3}$ predicted by HMF theory. Inset: Critical
    relaxation of the density of singly occupied vertices in a
    semi-log scale.  The horizontal line indicates the density
    $\rho=1/2$ predicted by the HMF theory, Eq.~\eqref{eq:rhok}.}
 \label{fig:rlx}
\end{figure}

The excellent agreement between HMF and QS simulations on annealed
networks is expected since the central MF hypotheses are fulfilled:
absence of dynamical correlations and a distance between nodes $\ell
\equiv 1$~\footnote{In an annealed substrate a given node interacts
  with any node of the network including itself due to the
  connectivity rewiring.} implying the exactness of the one-site
approximation. In real (quenched) networks, whose edges are frozen and
do not change in time, we usually still have the small-world property,
but dynamical correlations are unavoidably present. An additional
feature of TCP is that activity spreads through the backbone of
occupied vertices. So, a necessary condition for the maintenance of
activity is the existence of a giant component (GC) of connected
vertices in the backbone~\cite{newman2010}. This component is present
in our simulations as discussed in the end of this Section.

\begin{figure*}[t]
 \centering
 \includegraphics[width=15cm]{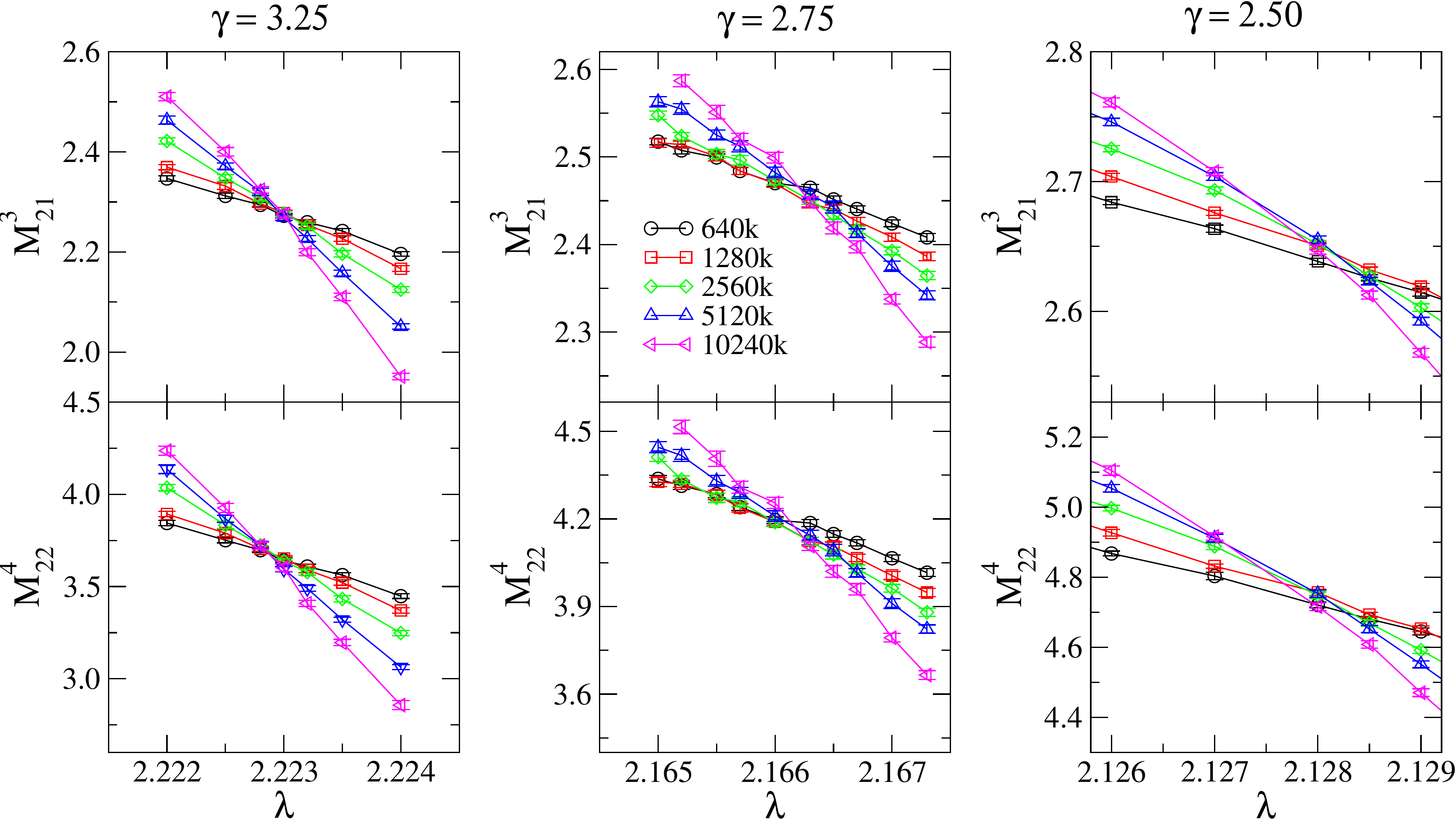}   
 % m2ndg275.pdf: 742x555 pixel, 72dpi, 26.18x19.58 cm, bb=0 0 742 555
 \caption{Moment ratios against creation rate of the TCP on quenched
   SF networks with different values of $\gamma$, for several system
   sizes. For the sake of comparison, all plots have the same
   horizontal range.}
 \label{fig:momquen}
\end{figure*}

% UCM networks~\cite{Catanzaro05} used in our simulations are
% free from degree-degree correlations and, then, suitable for a
% comparison with HMF theory developed in
% section~\ref{sec:mf}.

An important issue in the simulations is that, differently from the
standard CP on quenched networks~\cite{Ferreira_quenched}, the
relaxation of TCP in SF networks depends on initial
conditions. Figure~\ref{fig:rlx} shows the critical relaxation in a
quenched SF network ($\gamma=2.75$) for two initial conditions.  A
fully active initial condition leads to a metastable phase where the
system gets stuck into a pseudo-stationary state of very low
density. After a transient, which can be very long for large sizes,
the system evolves towards the real QS state.  However, if we start
with an initial condition having a small fraction of active vertices
and the remaining ones occupied by a single particle, the metastable
phase disappears and the system goes directly to the real QS state.
This dependence on the initial condition is easily understood by
considering that, in the early steps of a simulation with a fully
active initial condition, annihilation events are predominant due to
the lack of free space to create new particles. These annihilation
events fragment the backbone into several small components and
activity cannot spread until the backbone is regenerated, after
a very large transient.

We thus choose to start with a $2\%$ initial density of active
vertices in all simulations reported in the following.  Notice that
the density of singly occupied nodes goes quickly to values very close
to $1/2$ as shown in the inset of Fig.~\ref{fig:rlx}, but the final
quasi-stationary state takes much longer times in consonance with the
quasi-static approximation used in HMF calculations of
Sec.~\ref{sec:hmf}.

\begin{table}[b]
  \begin{center}
    \begin{tabular}{ccccc}
      \hline\hline
      $\gamma$ & $\lambda_c$ & $M_{11}^2$ & $M_{21}^3$  & $M_{22}^4$ \\ \hline
      2.50     & ~2.1280(5)~ & ~1.87(2)~  & ~2.65(2)~   & ~4.73(5)~  \\
      2.75     &  2.1661(2)  &  1.790(6)  &  2.43(3)    &  4.15(9)   \\
      3.25     &  2.2229(1)  &  1.712(8)  &  2.27(1)    &  3.71(4)   \\ 
      \hline \hline
    \end{tabular}
  \end{center}
  \caption{Critical points and moment ratios for TCP on quenched SF
    networks.} 
  \label{tab:critmom}
\end{table}

\begin{figure}
  \centering
  \includegraphics[width=8cm]{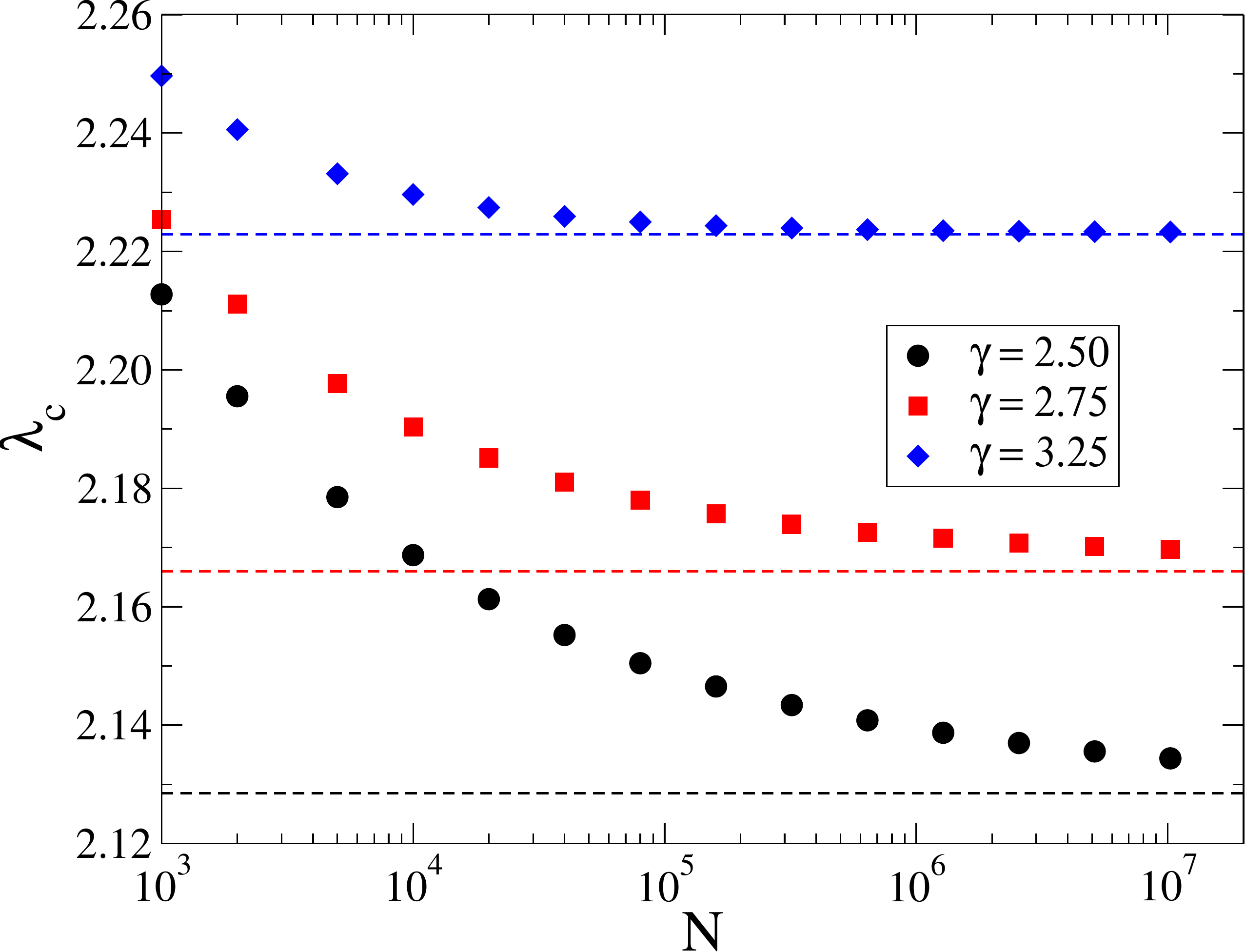}
  % lambdacHPA.pdf: 750x518 pixel, 72dpi, 26.46x18.27 cm, bb=0 0 750 518
  \caption{Comparison between critical points obtained in a
    homogeneous pair approximation (symbols) and QS simulations
    (lines) of the TCP on quenched SF networks. HPA results correspond
    to Eq.~(\ref{eq:lbc_pair}) with the coordination $q$ replaced by
    the average degree of the network.}
 \label{fig:HPA}
\end{figure}

Critical points were determined using the moments ratio method
described in section~\ref{sec:ann}. Figure~\ref{fig:momquen} shows
third and fourth order moments ratios against creation rates for
networks of different sizes and degree exponents. All curves intersect
closely at the same point yielding the critical points and moments
ratio values shown in Table~\ref{tab:critmom}.  These moments ratio
values are noticeably in good agreement with those obtained for CP
simulations on the same kind of
networks~\cite{Ferreira_quenched}. Notice also that the critical
points are larger than the HMF prediction $\lambda_c=2$, independently
of $\gamma$.  A satisfactory agreement between the
homogeneous pair approximation value for the critical point,
Eq.~\eqref{eq:lbc_pair}, and the QS simulations can be however
observed, see Fig.~\ref{fig:HPA}, again in conformity with the results
for the CP on quenched
networks~\cite{Ferreira_quenched,MunozPRL2010}. 
% In Fig.~\ref{fig:HPA} we compare HPA with critical points determined
% in simulations. It is known that the critical point determination
% becomes more precise when higher order cluster approximations
% (three-sites, four-sites, and so forth) are used in lattices models
% ~\cite{Avraham92}. However, the network heterogeneity must play an
% important role and the homogeneous theory is still an approximation
% even if we take higher orders clusters.  It is not surprising
% because HMF theory neglects all dynamical correlations.

\begin{figure}[t]
 \centering
 \includegraphics[width=8cm]{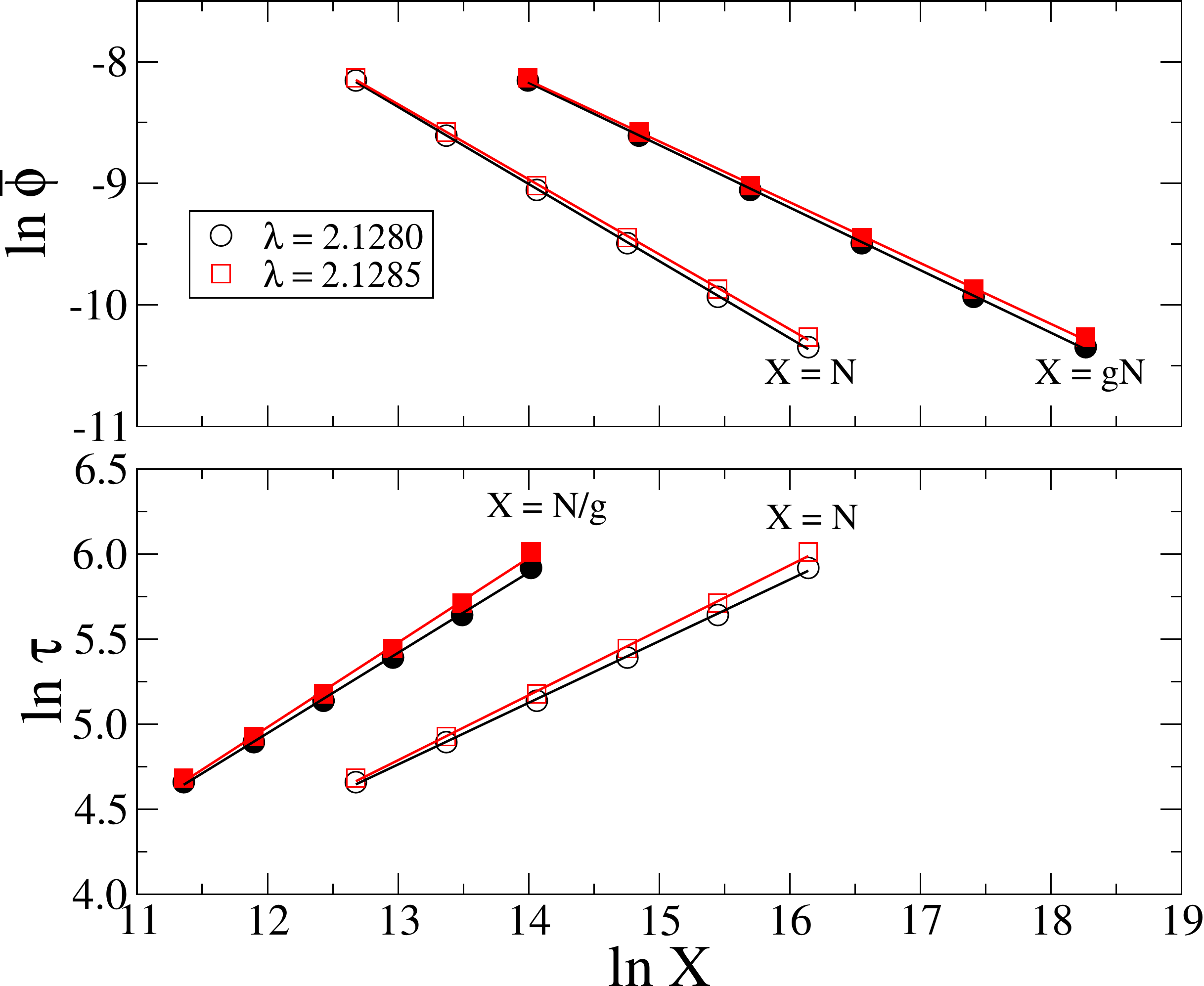}
 \caption{Finite size scaling of the QS density $\bar{\phi}$ and
   characteristic time $\tau$ around the critical point for quenched
   SF networks with $\gamma=2.50$. Solid lines are linear regressions
   used to estimate the scaling exponents. Open symbols refer to plots
   against size $N$ while filled symbols to the plots against size
   re-scaled by the factor $g = \ksq/\km^2$.}
 \label{fig:qscrit}
\end{figure}

Now in possession of accurate estimates of the critical points, we can
check the validity of HMF critical exponents. Figure~\ref{fig:qscrit}
shows the QS density and characteristic time for $\gamma=2.50$ around
the estimated critical point. Performing power law regression of these
quantities as a function of $N$, a very good agreement with HMF
exponents in Eq.~\eqref{eq:nu_alpha} is found, as shown in Table
\ref{tab:critexpo}. Using the full anomalous FSS predicted by HMF
theory, performing power law regressions of the form
$\bar{\phi}\sim(Ng)^{-S_{\nu}}$ and $\tau\sim (N/g)^{S_{\alpha}}$, all
numerically obtained exponents are in very good agreement with the HMF
result $S_\alpha=S_\nu=1/2$. The agreement of simulations with HMF
theory extends to all the values of $\gamma$ considered, as shown in
Table~\ref{tab:critexpo}.

\begin{table}[b]
  \begin{center}
    \begin{tabular}{ccccccc}
      \hline\hline
      $\gamma$&$\hat{\nu}$&$\hat{\nu}_{HMF}$&$S_\nu$&$\hat{\alpha}$&$\hat{\alpha}_{HMF}$&$S_\alpha$\\ \hline
      2.50    & ~0.62(1)~ &~0.625~          &~0.51(1)~  &~0.37(1)~ &~0.375~             &~0.48(1)~\\
      2.75    & ~0.57(1)~ &~0.5625~         &~0.496(8)~ &~0.42(1)~ &~0.4375~            &~0.50(1)~\\
      3.25    & ~0.50(1)~ &~0.500~          &~0.48(1)~  &~0.49(2)~ &~0.500~             &~0.51(2)~\\ 
      \hline \hline
    \end{tabular}
  \end{center}
  \caption{Critical exponents for TCP on quenched SF
    networks.  HMF exponents $\hat{\nu}_{HMF}$ and
    $\hat{\alpha}_{HMF}$, as given by Eq.~\eqref{eq:nu_alpha}, are included
    for comparison.}
\label{tab:critexpo}
\end{table}

\begin{figure}[t]
 \centering
 \includegraphics[width=8cm]{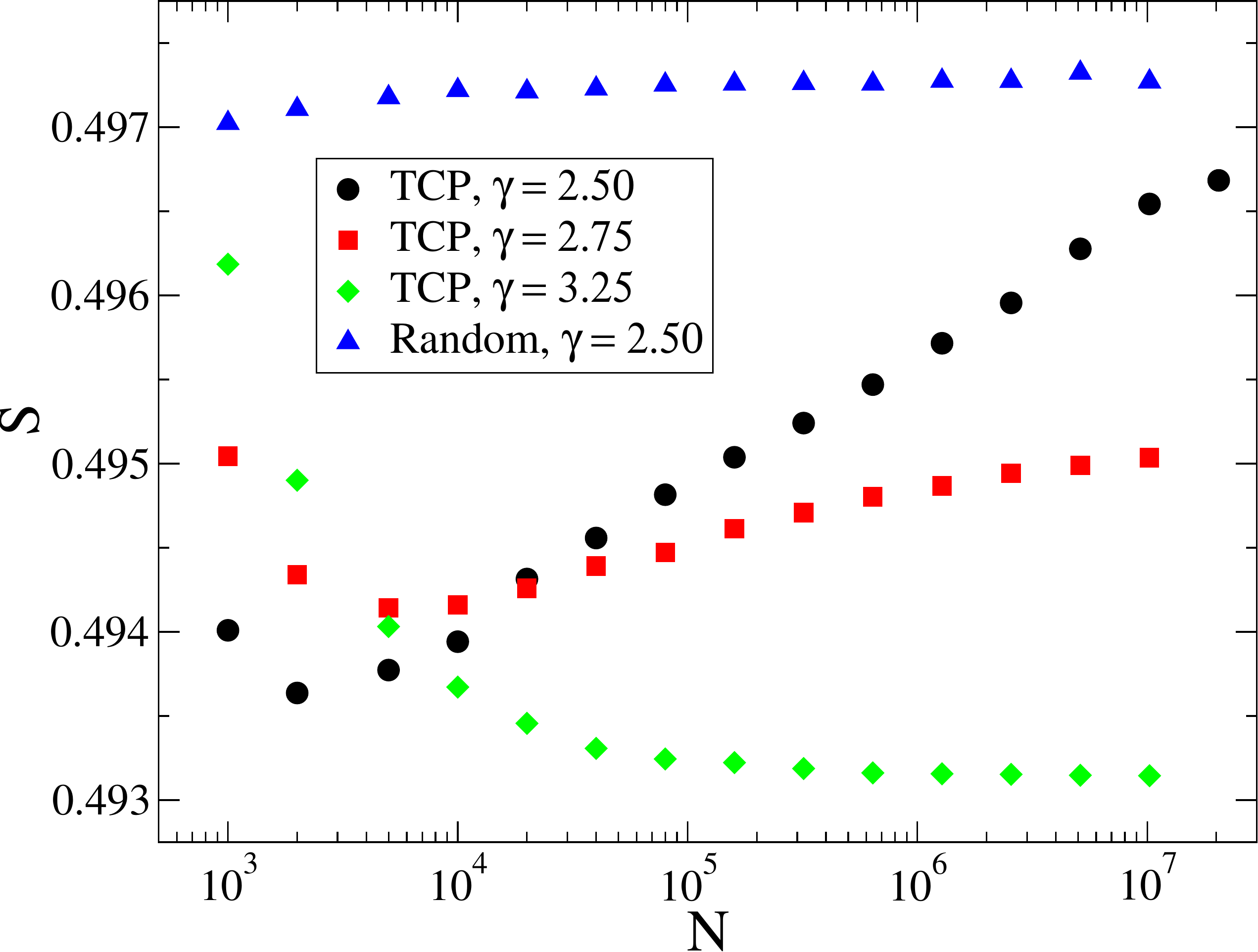}
 % giant.pdf: 750x562 pixel, 72dpi, 26.46x19.83 cm, bb=0 0 750 562
 \caption{Average size of the giant component against network size obtained 
 with the TCP dynamics or a random dilution of 50\% of the vertices. The random
dilution for other $\gamma$ values have the same behavior of $\gamma=2.50$.}
 \label{fig:giant}
\end{figure}

We have finally investigated the backbone of the network used for the
spreading of activity, which is defined as the largest connected
cluster (giant component) of vertices having at least one particle.
Figure~\ref{fig:giant} shows the average fraction of vertices
belonging to the backbone in the critical TCP against network
size. The size of the GC corresponding to a random dilution of 50\% is
also shown for the sake of comparison.  It is well known that SF
networks are resilient to random removal of nodes, such that a giant
component persists even if a finite fraction of the nodes is
removed~\cite{Cohen2000,PhysRevLett.85.5468}. A random dilution of
$50\%$ of the nodes in UCM networks leads to GCs containing almost the
totality of the non-diluted nodes (Fig.~\ref{fig:giant}). The critical
TCP dynamics generates a backbone with a similar fraction of
vertices. The second largest connected cluster contains just a few
nodes (typically 3 or 4 for a network size $N\simeq 10^7$) and,
therefore, the activity in regions disconnected from the backbone is
negligible.

Noticeably, while random dilution and critical TCP on homogeneous and
large networks ($\gamma=3.25$) generate GCs that do not vary with
network size, in the heterogeneous case we have a slow
(sub-logarithmic) increase, which is more apparent the smaller the
value of $\gamma$.  This dependence of the backbone size with $N$ is
however so slow (e.g., a relative variation of around $0.5\%$ when
size increases from $10^3$ to $10^7$ for $\gamma=2.50$) that its
effect is essentially negligible, as can be seen from the excellent
agreement with HMF theory shown by numerical simulations.

\section{Discussion}
\label{sec:conclu}

In the present work, we have studied the absorbing state phase
transition of a threshold contact process (TCP) in networks. The
model, possessing infinitely many absorbing configurations in the
thermodynamic limit, is analogous to the pair contact process
\cite{JensenPCP93} with a relaxation of the fermionic constraint: up
to two particles can occupy a vertex but only the particles in doubly
occupied vertices react (annihilate or create a new particle in a
nearest neighbor).

Applying the standard heterogeneous mean-field theory and extensive
large-scale numerical simulations based on the quasi-stationary state
method, we have shown that the critical behavior of the TCP is exactly
the same as the standard CP, and therefore that both models belong to
the so-called CP-HMF universality class in networks, which is defined,
at the FSS level, by the set of exponents given by
Eq.~\eqref{eq:nu_alpha}. At this respect, the apparent non mean-field
exponents reported in Ref.~\cite{da-yin10} for the related pair
contact process, a fermionic counterpart of the TCP, can probably be
attributed to the networks sizes considered in that work ($N \lesssim
10^4$), much smaller than the largest system sizes ($N\sim 10^7$) that
we attain here.

The CP-HMF universality class studied here can be conjectured to
include other CP-like models with a finite or infinite number of
absorbing states. However, its status is not as robust as in lattices,
since in networks microscopic details of the models can be enhanced
due to heterogeneity. For example, the
susceptible-infected-susceptible (SIS) epidemic model~\cite{RomuPRE01}
is a variation of the CP in which a vertex produces activity in all
its neighbors with the same strength, while in the CP the activity
production of a vertex is equally divided among all neighbors (factor
$1/k'$ in Eq.~(\ref{eq:Thetak})). According to the Janssen-Grassberger
conjecture~\cite{Janssen81,Grassberger82}, the universality class of
CP and SIS should be the same, and in fact both models belong to the
DP universality class in homogeneous lattices. In complex networks,
however, heterogeneity strongly affects both models and it renders the
SIS with critical exponents at the HMF level which are different from
the CP~\cite{RomuPRE01}. Moreover, the SIS can be shown to exhibit a
vanishing critical point in the thermodynamical limit if the maximal
connectivity is unbounded~\cite{RomuPRL2010,Ferreira12}. 

We can thus conjecture a splitting of universality classes, comprising
at least a CP-HMF and a SIS-HMF class, which are independent of the
number of absorbing states (as proved in the SIS case by the study of
the forest-fire model~\cite{Bancal10}) but depends on the way in which
activity spreads over nearest neighbors. Further work should be
devoted to the study of this universality class splitting in complex
networks.

\begin{acknowledgments}
  This work was partially supported by the Brazilian agencies CNPq,
  FAPEMIG, and CAPES.  R.P.-S.  acknowledges financial support from
  the Spanish MEC, under project No. FIS2010-21781-C02-01; the Junta
  de Andaluc\'{\i}a, under project No. P09-FQM4682; ICREA Academia,
  funded by the Generalitat de Catalunya; partial support by the NSF
  under Grant No. PHY1066293, and the hospitality of the Aspen Center
  for Physics, CO, USA, where part of this work was performed.
\end{acknowledgments}

\bibliographystyle{apsrev4-1}
%\bibliography{pcpnets}

\begin{thebibliography}{10}%
\makeatletter
\providecommand \@ifxundefined [1]{%
 \ifx #1\undefined \expandafter \@firstoftwo
 \else \expandafter \@secondoftwo
\fi
}%
\providecommand \@ifnum [1]{%
 \ifnum #1\expandafter \@firstoftwo
 \else \expandafter \@secondoftwo
\fi
}%
\providecommand \enquote [1]{``#1''}%
\providecommand \bibnamefont  [1]{#1}%
\providecommand \bibfnamefont [1]{#1}%
\providecommand \citenamefont [1]{#1}%
\providecommand\href[0]{\@sanitize\@href}%
\providecommand\@href[1]{\endgroup\@@startlink{#1}\endgroup\@@href}%
\providecommand\@@href[1]{#1\@@endlink}%
\providecommand \@sanitize [0]{\begingroup\catcode`\&12\catcode`\#12\relax}%
\@ifxundefined \pdfoutput {\@firstoftwo}{%
 \@ifnum{\z@=\pdfoutput}{\@firstoftwo}{\@secondoftwo}%
}{%
 \providecommand\@@startlink[1]{\leavevmode}%
 \providecommand\@@endlink[0]{}%
}{%
 \providecommand\@@startlink[1]{%
  \leavevmode
  \pdfstartlink
   attr{/Border[0 0 1 ]/H/I/C[0 1 1]}%
   user{/Subtype/Link/A<</Type/Action/S/URI/URI(#1)>>}%
  \relax
 }%
 \providecommand\@@endlink[0]{\pdfendlink}%
}%
\providecommand \url  [0]{\begingroup\@sanitize \@url }%
\providecommand \@url [1]{\endgroup\@href {#1}{\urlprefix}}%
\providecommand \urlprefix [0]{URL }%
\providecommand \Eprint[0]{\href }%
\@ifxundefined \urlstyle {%
  \providecommand \doi [1]{doi:\discretionary{}{}{}#1}%
}{%
  \providecommand \doi [0]{doi:\discretionary{}{}{}\begingroup
  \urlstyle{rm}\Url }%
}%
\providecommand \doibase [0]{http://dx.doi.org/}%
\providecommand \Doi[1]{\href{\doibase#1}}%
\providecommand \bibAnnote [3]{%
  \BibitemShut{#1}%
  \begin{quotation}\noindent
    \textsc{Key:}\ #2\\\textsc{Annotation:}\ #3%
  \end{quotation}%
}%
\providecommand \bibAnnoteFile [2]{%
  \IfFileExists{#2}{\bibAnnote {#1} {#2} {\input{#2}}}{}%
}%
\providecommand \typeout [0]{\immediate \write \m@ne }%
\providecommand \selectlanguage [0]{\@gobble}%
\providecommand \bibinfo [0]{\@secondoftwo}%
\providecommand \bibfield [0]{\@secondoftwo}%
\providecommand \translation [1]{[#1]}%
\providecommand \BibitemOpen[0]{}%
\providecommand \bibitemStop [0]{}%
\providecommand \bibitemNoStop [0]{.\EOS\space}%
\providecommand \EOS [0]{\spacefactor3000\relax}%
\providecommand \BibitemShut [1]{\csname bibitem#1\endcsname}%
%</preamble>
\bibitem{hugues95:_random_walks_II}%
  \BibitemOpen
  \bibfield{author}{%
  \bibinfo {author} {\bibfnamefont{B.~D.}\ \bibnamefont{Hugues}},\ }%
  \emph{\bibinfo {title} {Random walks and random environments}},\ Vol.\
  \bibinfo {volume} {II, Random Environments}\ (\bibinfo {publisher} {Clarendon
  Press},\ \bibinfo {address} {Oxford},\ \bibinfo {year} {1996})%
  \bibAnnoteFile{NoStop}{hugues95:_random_walks_II}%
\bibitem{havlindiffusiondisorder2000}%
  \BibitemOpen
  \bibfield{author}{%
  \bibinfo {author} {\bibfnamefont{D.}~\bibnamefont{Ben-Avraham}}\ and\
  \bibinfo {author} {\bibfnamefont{S.}~\bibnamefont{Havlin}},\ }%
  \emph{\bibinfo {title} {Diffusion and reactions in fractals and disordered
  systems}}\ (\bibinfo {publisher} {Cambridge University Press},\ \bibinfo
  {address} {Cambridge},\ \bibinfo {year} {2000})%
  \bibAnnoteFile{NoStop}{havlindiffusiondisorder2000}%
\bibitem{barabasi02}%
  \BibitemOpen
  \bibfield{author}{%
  \bibinfo {author} {\bibfnamefont{R.}~\bibnamefont{Albert}}\ and\ \bibinfo
  {author} {\bibfnamefont{A.-L.}\ \bibnamefont{Barab\'asi}},\ }%
  \bibfield{journal}{%
  \bibinfo {journal} {Rev. Mod. Phys.}\ }%
  \textbf{\bibinfo {volume} {74}},\ \bibinfo {pages} {47} (\bibinfo {year}
  {2002})%
  \bibAnnoteFile{NoStop}{barabasi02}%
\bibitem{mendesbook}%
  \BibitemOpen
  \bibfield{author}{%
  \bibinfo {author} {\bibfnamefont{S.~N.}\ \bibnamefont{Dorogovtsev}}\ and\
  \bibinfo {author} {\bibfnamefont{J.~F.~F.}\ \bibnamefont{Mendes}},\ }%
  \emph{\bibinfo {title} {Evolution of networks: From biological nets to the
  {I}nternet and {WWW}}}\ (\bibinfo {publisher} {Oxford University Press},\
  \bibinfo {address} {Oxford},\ \bibinfo {year} {2003})%
  \bibAnnoteFile{NoStop}{mendesbook}%
\bibitem{newman2010}%
  \BibitemOpen
  \bibfield{author}{%
  \bibinfo {author} {\bibfnamefont{M.}~\bibnamefont{Newman}},\ }%
  \emph{\bibinfo {title} {Networks: an introduction}}\ (\bibinfo {publisher}
  {Oxford University Press, Inc.},\ \bibinfo {year} {2010})%
  \bibAnnoteFile{NoStop}{newman2010}%
\bibitem{dorogovtsevRMP08}%
  \BibitemOpen
  \bibfield{author}{%
  \bibinfo {author} {\bibfnamefont{S.~N.}\ \bibnamefont{Dorogovtsev}}, \bibinfo
  {author} {\bibfnamefont{A.~V.}\ \bibnamefont{Goltsev}},\ and\ \bibinfo
  {author} {\bibfnamefont{J.~F.~F.}\ \bibnamefont{Mendes}},\ }%
  \bibfield{journal}{%
  \bibinfo {journal} {Rev. Mod. Phys.}\ }%
  \textbf{\bibinfo {volume} {80}},\ \bibinfo {pages} {1275} (\bibinfo {year}
  {2008})%
  \bibAnnoteFile{NoStop}{dorogovtsevRMP08}%
\bibitem{barratbook}%
  \BibitemOpen
  \bibfield{author}{%
  \bibinfo {author} {\bibfnamefont{A.}~\bibnamefont{Barrat}}, \bibinfo {author}
  {\bibfnamefont{M.}~\bibnamefont{Barth\'{e}lemy}},\ and\ \bibinfo {author}
  {\bibfnamefont{A.}~\bibnamefont{Vespignani}},\ }%
  \emph{\bibinfo {title} {Dynamical Processes on Complex Networks}}\ (\bibinfo
  {publisher} {Cambridge University Press},\ \bibinfo {address} {Cambridge},\
  \bibinfo {year} {2008})%
  \bibAnnoteFile{NoStop}{barratbook}%
\bibitem{pv01a}%
  \BibitemOpen
  \bibfield{author}{%
  \bibinfo {author} {\bibfnamefont{R.}~\bibnamefont{Pastor-Satorras}}\ and\
  \bibinfo {author} {\bibfnamefont{A.}~\bibnamefont{Vespignani}},\ }%
  \bibfield{journal}{%
  \bibinfo {journal} {Phys. Rev. Lett.}\ }%
  \textbf{\bibinfo {volume} {86}},\ \bibinfo {pages} {3200} (\bibinfo {year}
  {2001})%
  \bibAnnoteFile{NoStop}{pv01a}%
\bibitem{May01}%
  \BibitemOpen
  \bibfield{author}{%
  \bibinfo {author} {\bibfnamefont{R.~M.}\ \bibnamefont{May}}\ and\ \bibinfo
  {author} {\bibfnamefont{A.~L.}\ \bibnamefont{Lloyd}},\ }%
  \bibfield{journal}{%
  \bibinfo {journal} {Phys. Rev. E}\ }%
  \textbf{\bibinfo {volume} {64}},\ \bibinfo {pages} {066112} (\bibinfo {year}
  {2001})%
  \bibAnnoteFile{NoStop}{May01}%
\bibitem{PhysRevE.66.016128}%
  \BibitemOpen
  \bibfield{author}{%
  \bibinfo {author} {\bibfnamefont{M.~E.~J.}\ \bibnamefont{Newman}},\ }%
  \bibfield{journal}{%
  \Doi{10.1103/PhysRevE.66.016128}{\bibinfo {journal} {Phys. Rev. E}}\ }%
  \textbf{\bibinfo {volume} {66}},\ \bibinfo {pages} {016128} (\bibinfo {year}
  {2002})%
  \bibAnnoteFile{NoStop}{PhysRevE.66.016128}%
\bibitem{Cohen2000}%
  \BibitemOpen
  \bibfield{author}{%
  \bibinfo {author} {\bibfnamefont{R.}~\bibnamefont{Cohen}}, \bibinfo {author}
  {\bibfnamefont{K.}~\bibnamefont{Erez}}, \bibinfo {author}
  {\bibfnamefont{D.}~\bibnamefont{{ben-Avraham}}},\ and\ \bibinfo {author}
  {\bibfnamefont{S.}~\bibnamefont{Havlin}},\ }%
  \bibfield{journal}{%
  \bibinfo {journal} {Phys. Rev. Lett.}\ }%
  \textbf{\bibinfo {volume} {85}},\ \bibinfo {pages} {4626} (\bibinfo {year}
  {2000})%
  \bibAnnoteFile{NoStop}{Cohen2000}%
\bibitem{PhysRevLett.85.5468}%
  \BibitemOpen
  \bibfield{author}{%
  \bibinfo {author} {\bibfnamefont{D.~S.}\ \bibnamefont{Callaway}}, \bibinfo
  {author} {\bibfnamefont{M.~E.~J.}\ \bibnamefont{Newman}}, \bibinfo {author}
  {\bibfnamefont{S.~H.}\ \bibnamefont{Strogatz}},\ and\ \bibinfo {author}
  {\bibfnamefont{D.~J.}\ \bibnamefont{Watts}},\ }%
  \bibfield{journal}{%
  \bibinfo {journal} {Phys. Rev. Lett.}\ }%
  \textbf{\bibinfo {volume} {85}},\ \bibinfo {pages} {5468} (\bibinfo {year}
  {2000})%
  \bibAnnoteFile{NoStop}{PhysRevLett.85.5468}%
\bibitem{ArenasPhysRep2008}%
  \BibitemOpen
  \bibfield{author}{%
  \bibinfo {author} {\bibfnamefont{A.}~\bibnamefont{Arenas}}, \bibinfo {author}
  {\bibfnamefont{A.}~\bibnamefont{Diaz-Guilera}}, \bibinfo {author}
  {\bibfnamefont{J.}~\bibnamefont{Kurths}}, \bibinfo {author}
  {\bibfnamefont{Y.}~\bibnamefont{Moreno}},\ and\ \bibinfo {author}
  {\bibfnamefont{C.}~\bibnamefont{Zhou}},\ }%
  \bibfield{journal}{%
  \bibinfo {journal} {Phys. Rep.}\ }%
  \textbf{\bibinfo {volume} {469}},\ \bibinfo {pages} {93 } (\bibinfo {year}
  {2008})%
  \bibAnnoteFile{NoStop}{ArenasPhysRep2008}%
\bibitem{marro1999npt}%
  \BibitemOpen
  \bibfield{author}{%
  \bibinfo {author} {\bibfnamefont{J.}~\bibnamefont{Marro}}\ and\ \bibinfo
  {author} {\bibfnamefont{R.}~\bibnamefont{Dickman}},\ }%
  \emph{\bibinfo {title} {{Nonequilibrium Phase Transitions in Lattice
  Models}}}\ (\bibinfo {publisher} {Cambridge University Press},\ \bibinfo
  {address} {Cambridge},\ \bibinfo {year} {1999})%
  \bibAnnoteFile{NoStop}{marro1999npt}%
\bibitem{Henkel}%
  \BibitemOpen
  \bibfield{author}{%
  \bibinfo {author} {\bibfnamefont{M.}~\bibnamefont{Henkel}}, \bibinfo {author}
  {\bibfnamefont{H.}~\bibnamefont{Hinrichsen}},\ and\ \bibinfo {author}
  {\bibfnamefont{S.}~\bibnamefont{L\"ubeck}},\ }%
  \emph{\bibinfo {title} {Non-equilibrium phase transition: Absorbing Phase
  Transitions}}\ (\bibinfo {publisher} {Springer Verlag},\ \bibinfo {address}
  {Netherlands},\ \bibinfo {year} {2008})%
  \bibAnnoteFile{NoStop}{Henkel}%
\bibitem{harris74}%
  \BibitemOpen
  \bibfield{author}{%
  \bibinfo {author} {\bibfnamefont{T.~E.}\ \bibnamefont{Harris}},\ }%
  \bibfield{journal}{%
  \bibinfo {journal} {Ann. Prob.}\ }%
  \textbf{\bibinfo {volume} {2}},\ \bibinfo {pages} {969} (\bibinfo {year}
  {1974})%
  \bibAnnoteFile{NoStop}{harris74}%
\bibitem{RomuPRL2008}%
  \BibitemOpen
  \bibfield{author}{%
  \bibinfo {author} {\bibfnamefont{C.}~\bibnamefont{Castellano}}\ and\ \bibinfo
  {author} {\bibfnamefont{R.}~\bibnamefont{Pastor-Satorras}},\ }%
  \bibfield{journal}{%
  \bibinfo {journal} {Phys. Rev. Lett.}\ }%
  \textbf{\bibinfo {volume} {100}},\ \bibinfo {pages} {148701} (\bibinfo {year}
  {2008})%
  \bibAnnoteFile{NoStop}{RomuPRL2008}%
\bibitem{bogunaPRE2009}%
  \BibitemOpen
  \bibfield{author}{%
  \bibinfo {author} {\bibfnamefont{M.}~\bibnamefont{{Bogu\~{n}\'{a}}}},
  \bibinfo {author} {\bibfnamefont{C.}~\bibnamefont{Castellano}},\ and\
  \bibinfo {author} {\bibfnamefont{R.}~\bibnamefont{Pastor-Satorras}},\ }%
  \bibfield{journal}{%
  \bibinfo {journal} {Phys. Rev. E}\ }%
  \textbf{\bibinfo {volume} {79}},\ \bibinfo {pages} {036110} (\bibinfo {year}
  {2009})%
  \bibAnnoteFile{NoStop}{bogunaPRE2009}%
\bibitem{Ferreira_annealed}%
  \BibitemOpen
  \bibfield{author}{%
  \bibinfo {author} {\bibfnamefont{S.~C.}\ \bibnamefont{Ferreira}}, \bibinfo
  {author} {\bibfnamefont{R.~S.}\ \bibnamefont{Ferreira}},\ and\ \bibinfo
  {author} {\bibfnamefont{R.}~\bibnamefont{Pastor-Satorras}},\ }%
  \bibfield{journal}{%
  \bibinfo {journal} {Phys. Rev. E}\ }%
  \textbf{\bibinfo {volume} {83}},\ \bibinfo {pages} {066113} (\bibinfo {year}
  {2011})%
  \bibAnnoteFile{NoStop}{Ferreira_annealed}%
\bibitem{Ferreira_quenched}%
  \BibitemOpen
  \bibfield{author}{%
  \bibinfo {author} {\bibfnamefont{S.~C.}\ \bibnamefont{Ferreira}}, \bibinfo
  {author} {\bibfnamefont{R.~S.}\ \bibnamefont{Ferreira}}, \bibinfo {author}
  {\bibfnamefont{C.}~\bibnamefont{Castellano}},\ and\ \bibinfo {author}
  {\bibfnamefont{R.}~\bibnamefont{Pastor-Satorras}},\ }%
  \bibfield{journal}{%
  \bibinfo {journal} {Phys. Rev. E}\ }%
  \textbf{\bibinfo {volume} {84}},\ \bibinfo {pages} {066102} (\bibinfo {year}
  {2011})%
  \bibAnnoteFile{NoStop}{Ferreira_quenched}%
\bibitem{WIJLAND2003}%
  \BibitemOpen
  \bibfield{author}{%
  \bibinfo {author} {\bibfnamefont{F.}~\bibnamefont{{van Wijland}}},\ }%
  \bibfield{journal}{%
  \bibinfo {journal} {{Brazilian Journal of Physics}}\ }%
  \textbf{\bibinfo {volume} {33}},\ \bibinfo {pages} {551 } (\bibinfo {year}
  {2003})%
  \bibAnnoteFile{NoStop}{WIJLAND2003}%
\bibitem{Drossel92}%
  \BibitemOpen
  \bibfield{author}{%
  \bibinfo {author} {\bibfnamefont{B.}~\bibnamefont{Drossel}}\ and\ \bibinfo
  {author} {\bibfnamefont{F.}~\bibnamefont{Schwabl}},\ }%
  \bibfield{journal}{%
  \bibinfo {journal} {Phys. Rev. Lett.}\ }%
  \textbf{\bibinfo {volume} {69}},\ \bibinfo {pages} {1629} (\bibinfo {year}
  {1992})%
  \bibAnnoteFile{NoStop}{Drossel92}%
\bibitem{PhysRevE.57.5095}%
  \BibitemOpen
  \bibfield{author}{%
  \bibinfo {author} {\bibfnamefont{R.}~\bibnamefont{Dickman}}, \bibinfo
  {author} {\bibfnamefont{A.}~\bibnamefont{Vespignani}},\ and\ \bibinfo
  {author} {\bibfnamefont{S.}~\bibnamefont{Zapperi}},\ }%
  \bibfield{journal}{%
  \bibinfo {journal} {Phys. Rev. E}\ }%
  \textbf{\bibinfo {volume} {57}},\ \bibinfo {pages} {5095} (\bibinfo {year}
  {1998})%
  \bibAnnoteFile{NoStop}{PhysRevE.57.5095}%
\bibitem{JensenPCP93}%
  \BibitemOpen
  \bibfield{author}{%
  \bibinfo {author} {\bibfnamefont{I.}~\bibnamefont{Jensen}},\ }%
  \bibfield{journal}{%
  \bibinfo {journal} {Phys. Rev. Lett.}\ }%
  \textbf{\bibinfo {volume} {70}},\ \bibinfo {pages} {1465} (\bibinfo {year}
  {1993})%
  \bibAnnoteFile{NoStop}{JensenPCP93}%
\bibitem{Goh03}%
  \BibitemOpen
  \bibfield{author}{%
  \bibinfo {author} {\bibfnamefont{K.-I.}\ \bibnamefont{Goh}}, \bibinfo
  {author} {\bibfnamefont{D.-S.}\ \bibnamefont{Lee}}, \bibinfo {author}
  {\bibfnamefont{B.}~\bibnamefont{Kahng}},\ and\ \bibinfo {author}
  {\bibfnamefont{D.}~\bibnamefont{Kim}},\ }%
  \bibfield{journal}{%
  \bibinfo {journal} {Phys. Rev. Lett.}\ }%
  \textbf{\bibinfo {volume} {91}},\ \bibinfo {pages} {148701} (\bibinfo {year}
  {2003})%
  \bibAnnoteFile{NoStop}{Goh03}%
\bibitem{Andrade10}%
  \BibitemOpen
  \bibfield{author}{%
  \bibinfo {author} {\bibfnamefont{D.~O.}\ \bibnamefont{Cajueiro}}\ and\
  \bibinfo {author} {\bibfnamefont{R.}~\bibnamefont{Andrade}},\ }%
  \bibfield{journal}{%
  \bibinfo {journal} {Eur. Phys. Jour. B}\ }%
  \textbf{\bibinfo {volume} {77}},\ \bibinfo {pages} {291} (\bibinfo {year}
  {2010})%
  \bibAnnoteFile{NoStop}{Andrade10}%
\bibitem{Lee04}%
  \BibitemOpen
  \bibfield{author}{%
  \bibinfo {author} {\bibfnamefont{D.-S.}\ \bibnamefont{Lee}}, \bibinfo
  {author} {\bibfnamefont{K.-I.}\ \bibnamefont{Goh}}, \bibinfo {author}
  {\bibfnamefont{B.}~\bibnamefont{Kahng}},\ and\ \bibinfo {author}
  {\bibfnamefont{D.}~\bibnamefont{Kim}},\ }%
  \bibfield{journal}{%
  \bibinfo {journal} {Physica A}\ }%
  \textbf{\bibinfo {volume} {338}},\ \bibinfo {pages} {84 } (\bibinfo {year}
  {2004})%
  \bibAnnoteFile{NoStop}{Lee04}%
\bibitem{Bancal10}%
  \BibitemOpen
  \bibfield{author}{%
  \bibinfo {author} {\bibfnamefont{J.-D.}\ \bibnamefont{Bancal}}\ and\ \bibinfo
  {author} {\bibfnamefont{R.}~\bibnamefont{Pastor-Satorras}},\ }%
  \bibfield{journal}{%
  \bibinfo {journal} {Eur. Phys. Jour. B}\ }%
  \textbf{\bibinfo {volume} {76}},\ \bibinfo {pages} {109} (\bibinfo {year}
  {2010})%
  \bibAnnoteFile{NoStop}{Bancal10}%
\bibitem{da-yin10}%
  \BibitemOpen
  \bibfield{author}{%
  \bibinfo {author} {\bibfnamefont{H.}~\bibnamefont{Da-Yin}}\ and\ \bibinfo
  {author} {\bibfnamefont{W.}~\bibnamefont{Lie-Yan}},\ }%
  \bibfield{journal}{%
  \bibinfo {journal} {Chinese Phys. Lett.}\ }%
  \textbf{\bibinfo {volume} {27}},\ \bibinfo {pages} {098901} (\bibinfo {year}
  {2010})%
  \bibAnnoteFile{NoStop}{da-yin10}%
\bibitem{cardy88}%
  \BibitemOpen
  \bibinfo {editor} {\bibfnamefont{J.~L.}\ \bibnamefont{Cardy}},\ ed.,\
  \emph{\bibinfo {title} {Finite Size Scaling}},\ Vol.~\bibinfo {volume} {2}\
  (\bibinfo {publisher} {North Holland},\ \bibinfo {address} {Amsterdam},\
  \bibinfo {year} {1988})%
  \bibAnnoteFile{NoStop}{cardy88}%
\bibitem{Vespignani12}%
  \BibitemOpen
  \bibfield{author}{%
  \bibinfo {author} {\bibfnamefont{A.}~\bibnamefont{Vespignani}},\ }%
  \bibfield{journal}{%
  \Doi{10.1038/nphys2160}{\bibinfo {journal} {Nat. Phys.}}\ }%
  \textbf{\bibinfo {volume} {8}},\ \bibinfo {pages} {32} (\bibinfo {year}
  {2012})%
  \bibAnnoteFile{NoStop}{Vespignani12}%
\bibitem{Mancebo2005}%
  \BibitemOpen
  \bibfield{author}{%
  \bibinfo {author} {\bibfnamefont{M.~M.}\ \bibnamefont{de~Oliveira}}\ and\
  \bibinfo {author} {\bibfnamefont{R.}~\bibnamefont{Dickman}},\ }%
  \bibfield{journal}{%
  \bibinfo {journal} {Phys. Rev. E}\ }%
  \textbf{\bibinfo {volume} {71}},\ \bibinfo {pages} {016129} (\bibinfo {year}
  {2005})%
  \bibAnnoteFile{NoStop}{Mancebo2005}%
\bibitem{RomuPRL2006}%
  \BibitemOpen
  \bibfield{author}{%
  \bibinfo {author} {\bibfnamefont{C.}~\bibnamefont{Castellano}}\ and\ \bibinfo
  {author} {\bibfnamefont{R.}~\bibnamefont{Pastor-Satorras}},\ }%
  \bibfield{journal}{%
  \bibinfo {journal} {Phys. Rev. Lett.}\ }%
  \textbf{\bibinfo {volume} {96}},\ \bibinfo {pages} {038701} (\bibinfo {year}
  {2006})%
  \bibAnnoteFile{NoStop}{RomuPRL2006}%
\bibitem{alexei}%
  \BibitemOpen
  \bibfield{author}{%
  \bibinfo {author} {\bibfnamefont{R.}~\bibnamefont{Pastor-Satorras}}, \bibinfo
  {author} {\bibfnamefont{A.}~\bibnamefont{V{\'a}zquez}},\ and\ \bibinfo
  {author} {\bibfnamefont{A.}~\bibnamefont{Vespignani}},\ }%
  \bibfield{journal}{%
  \bibinfo {journal} {Phys. Rev. Lett.}\ }%
  \textbf{\bibinfo {volume} {87}},\ \bibinfo {pages} {258701} (\bibinfo {year}
  {2001})%
  \bibAnnoteFile{NoStop}{alexei}%
\bibitem{NohPRE2009}%
  \BibitemOpen
  \bibfield{author}{%
  \bibinfo {author} {\bibfnamefont{J.~D.}\ \bibnamefont{Noh}}\ and\ \bibinfo
  {author} {\bibfnamefont{H.}~\bibnamefont{Park}},\ }%
  \bibfield{journal}{%
  \bibinfo {journal} {Phys. Rev. E}\ }%
  \textbf{\bibinfo {volume} {79}},\ \bibinfo {pages} {056115} (\bibinfo {year}
  {2009})%
  \bibAnnoteFile{NoStop}{NohPRE2009}%
\bibitem{mariancutofss}%
  \BibitemOpen
  \bibfield{author}{%
  \bibinfo {author} {\bibfnamefont{M.}~\bibnamefont{{Bogu\~{n}\'{a}}}},
  \bibinfo {author} {\bibfnamefont{R.}~\bibnamefont{Pastor-Satorras}},\ and\
  \bibinfo {author} {\bibfnamefont{A.}~\bibnamefont{Vespignani}},\ }%
  \bibfield{journal}{%
  \bibinfo {journal} {Eur. Phys. J. B}\ }%
  \textbf{\bibinfo {volume} {38}},\ \bibinfo {pages} {205} (\bibinfo {year}
  {2004})%
  \bibAnnoteFile{NoStop}{mariancutofss}%
\bibitem{Kockelkoren}%
  \BibitemOpen
  \bibfield{author}{%
  \bibinfo {author} {\bibfnamefont{J.}~\bibnamefont{Kockelkoren}}\ and\
  \bibinfo {author} {\bibfnamefont{H.}~\bibnamefont{Chat\'e}},\ }%
  \bibfield{journal}{%
  \bibinfo {journal} {Phys. Rev. Lett.}\ }%
  \textbf{\bibinfo {volume} {90}},\ \bibinfo {pages} {125701} (\bibinfo {year}
  {2003})%
  \bibAnnoteFile{NoStop}{Kockelkoren}%
\bibitem{Avraham92}%
  \BibitemOpen
  \bibfield{author}{%
  \bibinfo {author} {\bibfnamefont{D.}~\bibnamefont{ben Avraham}}\ and\
  \bibinfo {author} {\bibfnamefont{J.}~\bibnamefont{K\"ohler}},\ }%
  \bibfield{journal}{%
  \bibinfo {journal} {Phys. Rev. A}\ }%
  \textbf{\bibinfo {volume} {45}},\ \bibinfo {pages} {8358} (\bibinfo {year}
  {1992})%
  \bibAnnoteFile{NoStop}{Avraham92}%
\bibitem{MunozPRL2010}%
  \BibitemOpen
  \bibfield{author}{%
  \bibinfo {author} {\bibfnamefont{M.~A.}\ \bibnamefont{Mu\~noz}}, \bibinfo
  {author} {\bibfnamefont{R.}~\bibnamefont{Juh\'asz}}, \bibinfo {author}
  {\bibfnamefont{C.}~\bibnamefont{Castellano}},\ and\ \bibinfo {author}
  {\bibfnamefont{G.}~\bibnamefont{\'Odor}},\ }%
  \bibfield{journal}{%
  \bibinfo {journal} {Phys. Rev. Lett.}\ }%
  \textbf{\bibinfo {volume} {105}},\ \bibinfo {pages} {128701} (\bibinfo {year}
  {2010})%
  \bibAnnoteFile{NoStop}{MunozPRL2010}%
\bibitem{michelediffusion}%
  \BibitemOpen
  \bibfield{author}{%
  \bibinfo {author} {\bibfnamefont{M.}~\bibnamefont{Catanzaro}}, \bibinfo
  {author} {\bibfnamefont{M.}~\bibnamefont{{Bogu\~{n}\'{a}}}},\ and\ \bibinfo
  {author} {\bibfnamefont{R.}~\bibnamefont{Pastor-Satorras}},\ }%
  \bibfield{journal}{%
  \bibinfo {journal} {Phys. Rev. E}\ }%
  \textbf{\bibinfo {volume} {71}},\ \bibinfo {pages} {056104} (\bibinfo {year}
  {2005})%
  \bibAnnoteFile{NoStop}{michelediffusion}%
\bibitem{Catanzaro05}%
  \BibitemOpen
  \bibfield{author}{%
  \bibinfo {author} {\bibfnamefont{M.}~\bibnamefont{Catanzaro}}, \bibinfo
  {author} {\bibfnamefont{M.}~\bibnamefont{Bogu\~n\'a}},\ and\ \bibinfo
  {author} {\bibfnamefont{R.}~\bibnamefont{Pastor-Satorras}},\ }%
  \bibfield{journal}{%
  \bibinfo {journal} {Phys. Rev. E}\ }%
  \textbf{\bibinfo {volume} {71}},\ \bibinfo {pages} {027103} (\bibinfo {year}
  {2005})%
  \bibAnnoteFile{NoStop}{Catanzaro05}%
\bibitem{Landaubook}%
  \BibitemOpen
  \bibfield{author}{%
  \bibinfo {author} {\bibfnamefont{D.}~\bibnamefont{Landau}}\ and\ \bibinfo
  {author} {\bibfnamefont{K.}~\bibnamefont{Binder}},\ }%
  \emph{\bibinfo {title} {A Guide to Monte Carlo Simulations in Statistical
  Physics}},\ \bibinfo {edition} {3rd}\ ed.\ (\bibinfo {publisher} {Cambridge
  University Press},\ \bibinfo {address} {New York, NY, USA},\ \bibinfo {year}
  {2009})%
  \bibAnnoteFile{NoStop}{Landaubook}%
\bibitem{stauffer_annealed2005}%
  \BibitemOpen
  \bibfield{author}{%
  \bibinfo {author} {\bibfnamefont{D.}~\bibnamefont{Stauffer}}\ and\ \bibinfo
  {author} {\bibfnamefont{M.}~\bibnamefont{Sahimi}},\ }%
  \bibfield{journal}{%
  \bibinfo {journal} {Phys. Rev. E}\ }%
  \textbf{\bibinfo {volume} {72}},\ \bibinfo {pages} {46128} (\bibinfo {year}
  {2005})%
  \bibAnnoteFile{NoStop}{stauffer_annealed2005}%
\bibitem{RomuPRE01}%
  \BibitemOpen
  \bibfield{author}{%
  \bibinfo {author} {\bibfnamefont{R.}~\bibnamefont{Pastor-Satorras}}\ and\
  \bibinfo {author} {\bibfnamefont{A.}~\bibnamefont{Vespignani}},\ }%
  \bibfield{journal}{%
  \bibinfo {journal} {Phys. Rev. E}\ }%
  \textbf{\bibinfo {volume} {63}},\ \bibinfo {pages} {066117} (\bibinfo {year}
  {2001})%
  \bibAnnoteFile{NoStop}{RomuPRE01}%
\bibitem{Janssen81}%
  \BibitemOpen
  \bibfield{author}{%
  \bibinfo {author} {\bibfnamefont{H.~K.}\ \bibnamefont{Janssen}},\ }%
  \bibfield{journal}{%
  \bibinfo {journal} {Z. Phys. B: Condens. Matter}\ }%
  \textbf{\bibinfo {volume} {42}} (\bibinfo {year} {1981})%
  \bibAnnoteFile{NoStop}{Janssen81}%
\bibitem{Grassberger82}%
  \BibitemOpen
  \bibfield{author}{%
  \bibinfo {author} {\bibfnamefont{P.}~\bibnamefont{Grassberger}},\ }%
  \bibfield{journal}{%
  \bibinfo {journal} {Z. Phys. B Condens. Matter}\ }%
  \textbf{\bibinfo {volume} {47}},\ \bibinfo {pages} {365} (\bibinfo {year}
  {1982})%
  \bibAnnoteFile{NoStop}{Grassberger82}%
\bibitem{RomuPRL2010}%
  \BibitemOpen
  \bibfield{author}{%
  \bibinfo {author} {\bibfnamefont{C.}~\bibnamefont{Castellano}}\ and\ \bibinfo
  {author} {\bibfnamefont{R.}~\bibnamefont{Pastor-Satorras}},\ }%
  \bibfield{journal}{%
  \bibinfo {journal} {Phys. Rev. Lett.}\ }%
  \textbf{\bibinfo {volume} {105}},\ \bibinfo {pages} {218701} (\bibinfo {year}
  {2010})%
  \bibAnnoteFile{NoStop}{RomuPRL2010}%
\bibitem{Ferreira12}%
  \BibitemOpen
  \bibfield{author}{%
  \bibinfo {author} {\bibfnamefont{S.~C.}\ \bibnamefont{Ferreira}}, \bibinfo
  {author} {\bibfnamefont{C.}~\bibnamefont{Castellano}},\ and\ \bibinfo
  {author} {\bibfnamefont{R.}~\bibnamefont{Pastor-Satorras}},\ }%
  \bibfield{journal}{%
  \bibinfo {journal} {Phys. Rev. E}\ }
  \textbf{\bibinfo {volume} {86}},\ \bibinfo {pages} {041125}
   (\bibinfo {year} {2012})%
  \bibAnnoteFile{NoStop}{Ferreira12}%
\end{thebibliography}

\end{document}